\documentclass[aps,pra,amsmath,amssymb,twocolumn,groupedaddress,showpacs,showkeys]{revtex4-1}

\usepackage{latexsym}
\usepackage{amssymb}
\usepackage{amsmath}
\usepackage{graphicx}
\usepackage{dcolumn}
\usepackage{bm}
\usepackage{gensymb}
\usepackage[dvipsnames]{xcolor}
\usepackage[normalem]{ulem}

\begin{document}

\title{A trapped atom interferometer with ultracold Sr atoms}

\author{Xian Zhang}
\email{also at: The Abdus Salam International Centre for
Theoretical Physics (ICTP), Strada Costiera 11 34014
Trieste, Italy}
\author{Ruben Pablo del Aguila}
\author{Tommaso Mazzoni}
\author{Nicola Poli}
\author{Guglielmo M. Tino}
\email{Guglielmo.Tino@fi.infn.it}
\altaffiliation{- also at: CNR - Institute of Applied Physics "Nello Carrara", Via Madonna del Piano 10, Sesto Fiorentino, Italy}

\affiliation{Dipartimento di Fisica e Astronomia and LENS -
Universit\`{a} di Firenze, INFN - Sezione di Firenze, Via Sansone
1, 50019 Sesto Fiorentino, Italy}

\date{\today}

\begin{abstract}
We report on a trapped atom interferometer based on Bragg
diffraction and Bloch oscillations with alkaline-earth-metal atoms. We
use a Ramsey-Bord\'e Bragg interferometer with
\textsuperscript{88}Sr atoms combined with Bloch oscillations to
extend the interferometer time. Thanks to a long coherence time for
Bloch oscillations of \textsuperscript{88}Sr atoms, we observed
interference up to $1$ s evolution time in the lattice.
A detailed study of decoherence sources during the Bloch phase is
also presented. While still limited in sensitivity by lattice
lifetime and beam inhomogeneity this result opens the way to high
contrast trapped interferometers with extended interrogation
time.
\end{abstract}

\maketitle


\section{Introduction}
Due to their high sensitivities and accuracy, next generation
atom interferometers are the focus of study in several laboratories around
the world \cite{tino2013atom}. Research in the field has centered
on increasing the interferometric path, in order to vastly improve
the sensitivity of these devices. Recent advances in atom interferometer sensitivity
pave the way for these devices to be used in precision tests of
fundamental physical theories such as quantum theory
\cite{Gillot2013,Kovachy2015,Lopes2015,Manning2015}, quantum gravity
\cite{Amelino-Camelia2009} and gravitation \cite{Biedermann2015,Hartwig2015,Zhou2015,Duan2016},
determination of physical constants \cite{Bouchendira2011,Rosi2014},
and, eventually, observation 
of gravitational waves in the low frequency regime
\cite{tino2007p,Dimopoulos2009,Hohensee2011,Graham2013}.

The most direct way of increasing the sensitivity of atom interferometers
is to simply increase the interferometer time $T$. For vertical atom
interferometers, this can be achieved by using long vacuum tubes
to increase the free-fall time \cite{Kovachy2015,Zhou2011,Hartwig2015}.
Provided that atomic samples are sufficiently cold and there are no limitations
due to laser beam geometry, for a 10 m vacuum tube the interferometer time
$T$ can be $>$ 1s. However, $T$ is proportional to the
square root of the tube length and a much more
efficient way of extending $T$ is by trapping the atoms
against gravity during the interferometer sequence itself. 

While the majority of the atom interferometers considered 
rely on Raman transitions in alkali atoms, recent experiments have 
demonstrated the feasibility of performing Bragg interferometry with 
ultracold alkaline-earth-metal(-like) atoms \cite{Mazzoni2015,Jamison2014}.
Furthermore, long coherence time of Bloch oscillations has been observed in 
$^{88}$Sr \cite{Poli2011,Tarallo2014}. Hence, strontium is a possible candidate for 
interferometric schemes that take advantage of the long coherence time of Bloch 
oscillations in order to extend the interferometer time $T$ by trapping the 
atoms in a 1-D vertical optical lattice, providing an alternative
avenue towards dramatically increasing the sensitivity of atom interferometers
without the need for large atomic fountains.

As well as extending the interferometer time through coherent Bloch
oscillations, the lattice can act as a waveguide upon the atomic samples,
limiting their radial (horizontal) expansion
\cite{McDonald2013}. Radial expansion of the atomic cloud is one
of the principal limitations in interferometer time (with a given Raman
or Bragg laser beam size and inhomogeneity), and is a cause 
of interferometer contrast loss.

We demonstrate a high contrast atom interferometer based on 
a Ramsey-Bord\'e Bragg interferometer, where the interferometer time 
is extended with a long Bloch oscillation phase, up to 1 s. 
Similar schemes applied to Rb atoms have been presented in
\cite{Charriere2012,Andia2013}. Here, we extend the Bloch oscillation time
with respect to previous work by using a different atom, namely
$^{88}$Sr. The motivation for using strontium atoms for such an
interferometer is due to their unique properties: they have zero
nuclear spin in the ground state which results in a low sensitivity 
to first order Zeeman shifts due to stray magnetic fields, 
as well as a low collisional cross section at
low temperatures, ensuring a coherence time of Bloch oscillations $>100$ s
\cite{Tarallo2012} in vertical optical lattices. 
Selecting strontium allows conception of
high sensitivity atom interferometers with total time $T$ practically
unreachable through other means.

\section{Experimental Apparatus}
 
\begin{figure}[t]\begin{center}
\includegraphics[width=0.47\textwidth]{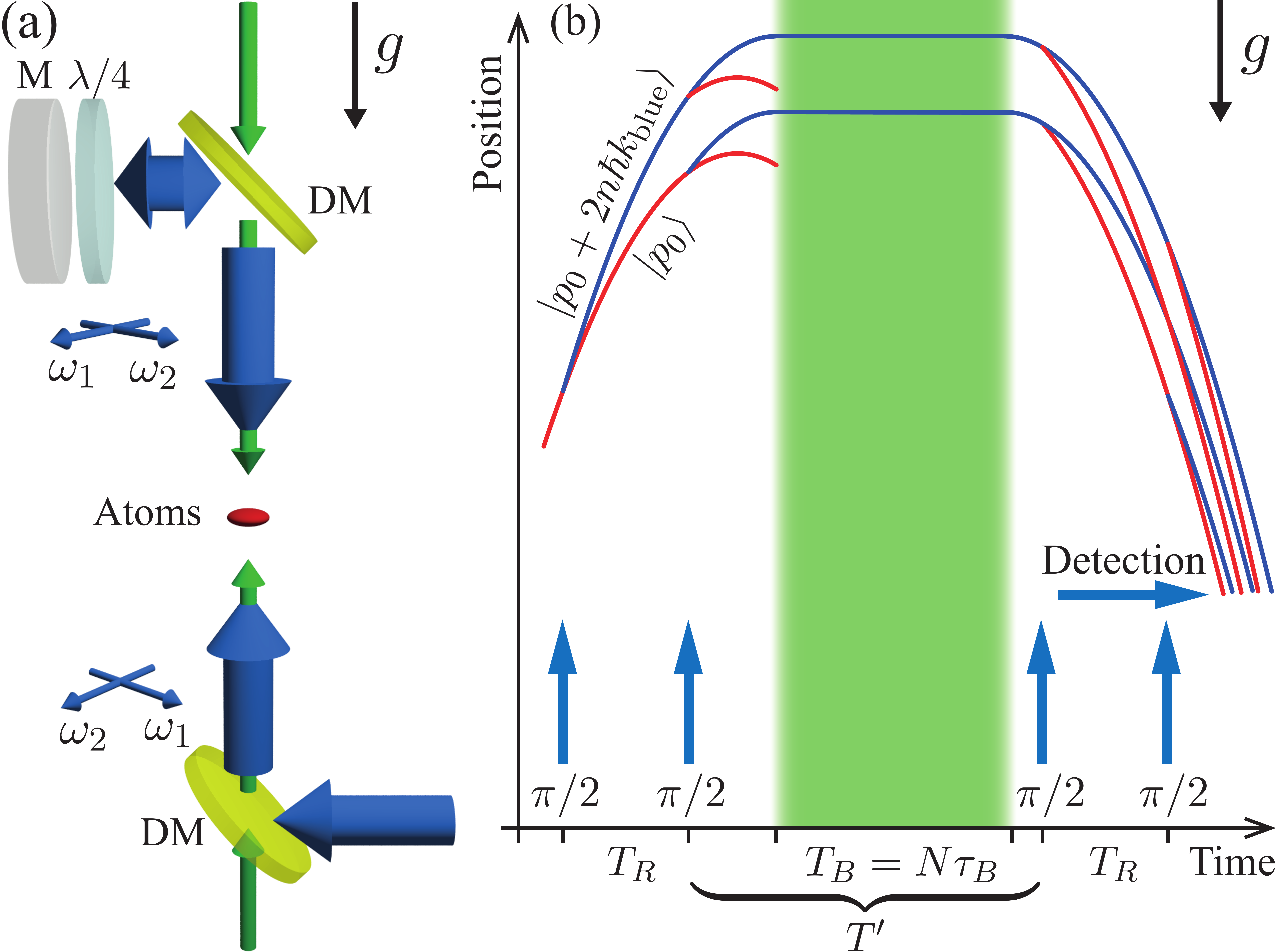}
\caption{(a) Schematic view of the experimental apparatus. Pre-cooled
\textsuperscript{88}Sr atoms trapped in a vertical optical lattice
operating at 532 nm (narrow green arrows) are launched upwards. Bragg pulses at 461 nm (thick blue arrows)
are then applied to first velocity select the atoms along the vertical
direction and subsequently to apply the interferometer sequence. 
The Bragg beams, with frequency $\omega_1$ and $\omega_2$ and orthogonal
polarizations, are sent from the bottom, rotated by a $\lambda/4$-wave plate and
retro-reflected by a mirror (M) on top. 
The lattice beams come from two independent fibers with linear polarizations 
and are superimposed onto the Bragg beams' center 
by means of two dichroic mirrors (DM).
(b) Atomic trajectories separated by $2n\hbar k_\mathrm{blue}$ in 
the interferometer. In the middle of
the interferometric sequence, the atoms are trapped in the green
vertical optical lattice (green shaded region) where they undergo Bloch oscillations.
}
\label{fig.apparatus}
\end{center}
\end{figure}

A schematic view of the experimental apparatus is shown in
Fig.~\ref{fig.apparatus}a.
In brief, a blue 461 nm laser red detuned by $\Delta=-10$ ~GHz from
the strong $^{1}S_{0}$--$^{1}P_{1}$ transition is employed for
inducing Bragg transitions. The Bragg laser source is similar to
the one described in \cite{Mazzoni2015} with an output power of
about 200 mW. The output is separated into two beams needed to drive
the Bragg transitions, and their intensity and frequency are 
independently controlled by two acousto-optical modulators (AOMs) 
placed between the laser source and the atomic sample. 
The Bragg frequency detuning is set to be
$\delta= 2\pi(f_{1}-f_{2})$, where $f_{1}$ and $f_{2}$ are the
driving radio frequencies (RF) of each AOM. The two
beams are coupled into a polarization-maintaining fiber with
mutually orthogonal polarizations; at the fiber output, a set of
telescopes collimate the beam with a $1/e^2$ radius of 3 mm. The
typical peak intensity is about $200$ mW/cm$^2$ for each beam.

The Bragg laser beams are aligned from below onto the atomic sample, and the
verticality of the beams is verified to within 1 mrad by means of a 
water surface reflection. The beams are then retro-reflected by a mirror
positioned atop the system. The reflecting mirror is not vibration-isolated due to
geometric limitations of the system. A quarter-wave plate is placed
before the retro-reflecting mirror to rotate the polarizations of
each reflected Bragg beam by 90\degree, forming two travelling waves
in opposite directions with which to induce the Bragg transitions. 
Each $n^{th}$ order Bragg transition transfers $2n\hbar k_\mathrm{blue}$ to every atom, 
where $k_\mathrm{blue}$ is the wave-vector of the 461 nm Bragg laser.

The Bragg AOMs are driven by a two-channel RF generator,
phase-locked to a 10 MHz Rubidium clock. The Bragg pulses are
generated and shaped into Gaussian profiles by a second
arbitrary-wave function generator mixed onto both channels of the
first RF generator, which serves as a variable attenuator for the
driving RF signals. In this text, pulse lengths are always
given in terms of the corresponding Gaussian $1/e^2$ width $\sigma$. 
A phase modulated RF source is mixed with one channel of the first RF 
generator to provide a phase shift for the interferometer.

A second laser, superimposed onto the path of the blue Bragg
laser, is used to provide the 1-D lattice for the Bloch evolution. 
The source laser for the lattice is a 
Coherent Verdi-V6 which delivers 6 W single mode
radiation at 532~nm. The green laser output is split equally into two beams,
with each beam passing through an AOM to stabilize the lattice intensity and
to control the frequency difference between the two beams. The two
lattice AOMs are driven by two RF synthesizers at $\sim100$~MHz. 
Two PID controllers modulate the RF synthesizers for
lattice intensity stabilization. A triggering signal with a time
constant of about 230~$\mu$s is summed onto the control signals
for adiabatic loading of the atoms into the lattice. The two lattice beams are then
each coupled into independent polarization-maintaining fibers. 
The two fiber outputs (about 1 W each) are collimated and sent from opposite
directions to the atoms, with the same linear polarization. 
Both lattice beams share the same vertical path as the Bragg beams, 
onto which they are superimposed by means of two dichroic mirrors placed 
below and above the vacuum chamber. 
At the position of the atoms the waist of each beam
is chosen to be 800~$\mu$m, with a Rayleigh length of about 3~m.
The lattice potential depth in this condition is about
$2E_{r,\mathrm{green}}$, where $E_{r,\mathrm{green}}=\hbar^2 k_\mathrm{green}^2/2m$ 
is the recoil energy of the green lattice, and $k_\mathrm{green}$ the wave-vector.

\section{Experimental sequence}
The cold atom preparation stage, as described in \cite{Mazzoni2015}, 
results in about $4\times10^6$ $^{88}$Sr atoms which are trapped
and cooled down to $1.2~ \mu$K with a two-stage magneto-optical
trap (MOT). The final radial (vertical) atomic distribution has a
full width at half maximum (FWHM) of $170$ $\mu$m (70 $\mu$m).

About 10\% of atoms are loaded directly from the MOT
into the lattice in order to give them an initial vertical launch. 
During the loading, an additional cooling stage occurs by setting 
the second-stage-MOT laser frequency closer to resonance 
and by reducing the MOT beam intensity. 
This additional cooling stage reduces the losses due to initial
evaporation of hot atoms loaded in the lattice and, as a result, 
there is an almost twofold increase in the number of atoms 
available for the interferometer. In these conditions, the atoms
loaded in the lattice reach a temperature of $\sim400$~nK, with a
typical lifetime of about 1~s, limited mainly by the background
vacuum.

Once loaded into the lattice, in order to do state preparation and 
to gain sufficient interferometer time,  
a vertical atom acceleration is achieved by chirping
the frequency detuning between the two lattice beams 
from 0~kHz to 850~kHz with a constant rate of 10~kHz/ms, 
corresponding to a constant acceleration of 2.59 m/s$^2$. 
The atomic cloud is then further
elevated for 40 ms in a moving lattice with constant velocity of
0.22~m/s, to a height of about 0.9~cm above the MOT position.
During this stage the atoms are adiabatically following the moving lattice 
in the first band; losses (about 20\% of the initial population) are mainly 
due to Landau-Zener tunneling in the acceleration phase. 
At this point, the remaining atoms are released from the
trap and a sequence of
three first order Bragg $\pi$ pulses is applied, each with 
an increasing duration (9 $\mu$s, 20 $\mu$s and 40 $\mu$s respectively).
The last pulse ensures the required velocity selection for the interferometer,
while the first two pulses are necessary to spatially separate the velocity 
selected atomic cloud from the residual atoms.
The efficiency of the velocity selection is about 25\%, leaving
about $4\times10^4$ atoms, with a typical vertical momentum width
of $\sim 0.15\,\hbar k_\mathrm{blue}$. 
For Bragg interferometers a narrow selection of the atomic
momentum is very important, not only for ensuring the high
coherence of initial atomic wave packet (which is necessary for a high
contrast interferometer signal \cite{Szigeti2012}), but also for
independent detection of the different interferometer channels.
In stark contrast to Raman inteferometers, the quantum interferometer 
states of a Bragg interferometer are encoded only onto the atoms'
external degrees of freedom, which are all detected by using 
standard on-resonance fluorescence/absorption detection. 
In this case, a narrow momentum distribution is
important in order to separate the interferometer output signals
after a reasonable time of flight.

The $n^{th}$ order Bragg diffraction has an effective Rabi
frequency
$\Omega_\mathrm{eff}=\Omega^n/[(8\omega_r)^{n-1}(n-1)!^2]$ \cite{Giltner1995,Muller2008},
where $\omega_r=\hbar k_\mathrm{blue}^2/2m=2\pi\times 10.7$~kHz is the
recoil frequency for the chosen Bragg laser and $\Omega$ is the two-photon Rabi frequency. In our experiment, the typical
Rabi frequency of a first order Bragg diffraction is
$\Omega_\mathrm{eff}=2\pi\times 20$~kHz for velocity selection pulses and
$\Omega_\mathrm{eff}=2\pi\times 80$~kHz for each of the interferometer's 
beam splitter pulses. The Fourier width of the beam splitter pulses is then
sufficiently larger than the atomic momentum distribution, so
most of the selected atoms are addressed for the interferometer.
The $\pi$ pulse efficiency is 95\% for first order Bragg
transitions and 80\% for second order transitions.

As shown in Fig.~\ref{fig.apparatus}(b), the interferometer consists
of two pairs of $10$ $\mu$s long $\pi/2$ pulses separated by a
time $T_{R}$ and a Bloch oscillation phase lasting for a time
$T_B=N\tau_B$, where $\tau_B= 2h/mg\lambda_\mathrm{green}$= 1.74 ms is the Bloch
period and $N$ the number of Bloch oscillations. While the first
pair of $\pi/2$ pulses are applied as the atomic cloud is
travelling upward (immediately after the velocity-selective pulses), the
two closing pairs are applied after the release from the lattice.

After the first $\pi/2$ pulse, the atomic wave packet is coherently split
into two states, one with a vertical momentum of $|p_0 \rangle$ and the other
with a vertical momentum of $|p_0+2n \hbar k_\mathrm{blue} \rangle$. 
The second $\pi/2$ pulse coherently splits each state again, resulting 
in four states, two with momentum $|p_0 \rangle$ and two with momentum
$|p_0+2n \hbar k_\mathrm{blue} \rangle$. When the atoms in the higher momentum
state reach the apogee, the green lattice beams are switched on
over 230 $\mu$s to adiabatically load the atoms into the first lattice band.
Meanwhile, the atoms in the lower momentum state free-fall away and do not
interfere with the trapped atoms (see Fig.~\ref{fig.apparatus}b).
The recapture efficiency of the atoms in the optical lattice after
the free-flight period mostly depends on the free fall expansion
of the atomic cloud after the launch that determines the final
size of the cloud with respect to the lattice beam waist (and
potential trap depth). In order to ensure a high recapture efficiency, 
the final launch velocity is set in order to limit the free-fall time 
between the second $\pi/2$ pulse and the switching-on of the lattice recapture 
to $\sim 20$ ms. With this choice, we typically obtain a recapture
efficiency of $\sim 90$\%. The timing of the Bragg pulses is also 
critical in order to avoid double Bragg diffraction \cite{Giese2013}; 
we set them to occur at least 5 ms away from the apogee time. 
Based on our launch and Bragg pulse timing parameters, 
the total time between two $\pi/2$ pulse pairs is $T'=
(38.8$~ms $-T_{R}) + N\tau_{B}$.

At the end of the interferometer sequence, fluorescence detection is
performed after 50 ms of time of flight (TOF).
We employ a linearly polarized probe beam, resonant with the
$^{1}S_{0}$--$^{1}P_{1}$ transition, placed $1.3$ cm below the MOT
position and retro-reflected to increase the intensity
of the fluorescence signal. To improve selectivity of different
momentum states, the probe beam is collimated into a sheet of
light with a vertical (horizontal) width of about 200~$\mu$m
(2~mm).

\section{Interferometer results}

\begin{figure}\begin{center}
\includegraphics[width=0.47 \textwidth]{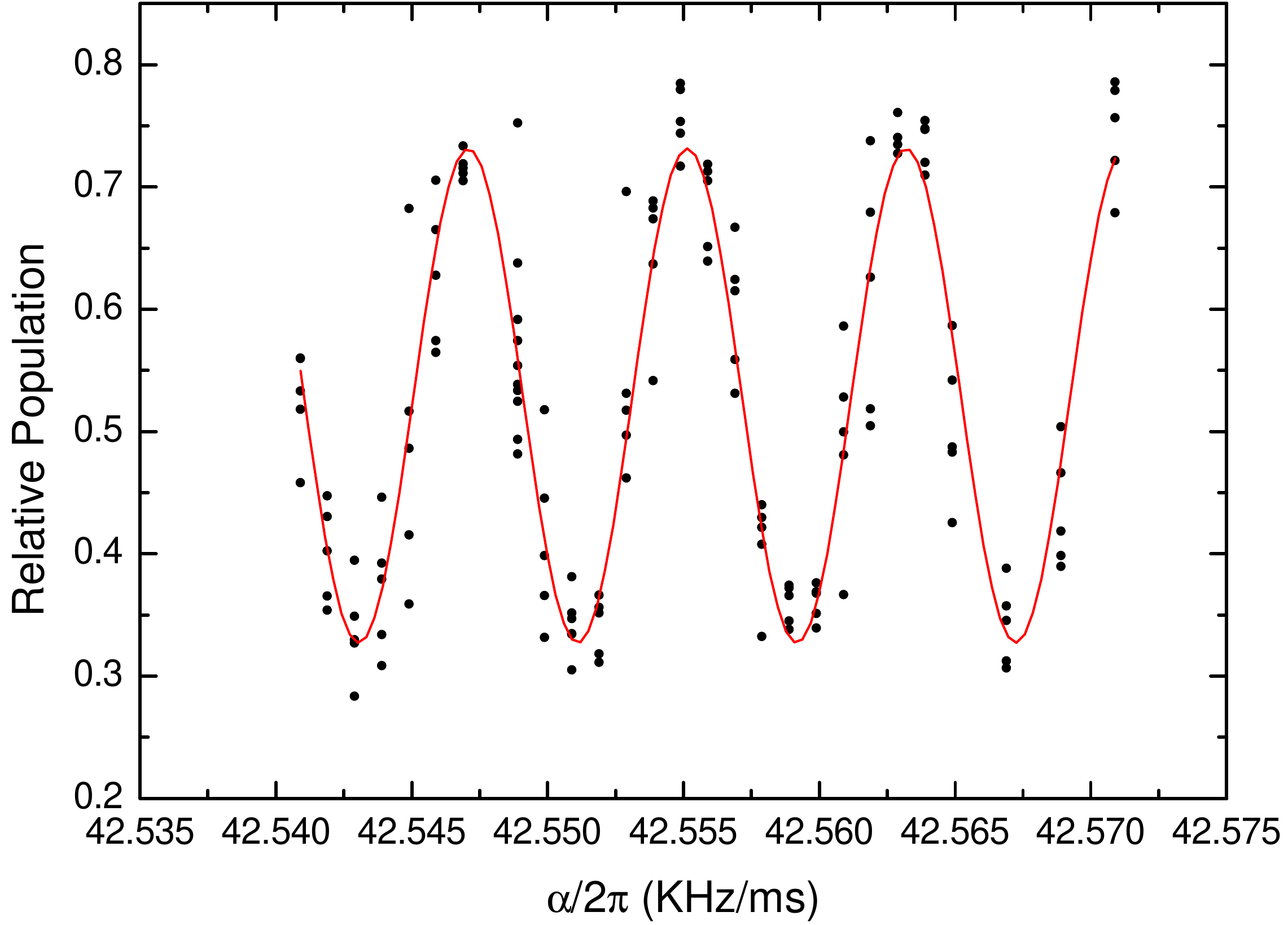}
\caption{Typical Ramsey-Bord\'e + Bloch interferometer fringes
for an order of Bragg transition $n=1$,
obtained by scanning the frequency chirping $\alpha$. In this
particular measurement we choose $T_{R} = 1$~ms and $T' = 123$~ms,
where $T_{B} = 85.2$~ms for $N=49$. The sinusoidal fit gives a
visibility of $0.4$.} \label{fig.fringes}
\end{center}
\end{figure}

In this section, we focus on the interferometer results,
especially the fringe contrast and its decay. The interferometer
output is recorded by measuring the relative population $P$ of the
momentum state $|p_0\rangle$ over the sum population of both states
$|p_0\rangle$ and $|p_0+2n \hbar k_\mathrm{blue} \rangle$:
$P=N_{|p_0\rangle}/(N_{|p_0\rangle}+N_{|p_0+2n\hbar k_\mathrm{blue}
\rangle})$. The interferometer fringes can be written as:
\begin{equation}
P(\phi)=P_{0}+\frac{C}{2} \cos(\phi)
\end{equation}
where P$_0\sim 0.5$ is a fringe offset and $C$ is the contrast.
The phase $\phi$ for the trapped Ramsey-Bord\'e + Bloch
interferometer is:
\begin{equation}
\phi=n(2k_\mathrm{blue}g-\alpha)T_{R}(T_{R}+T')
+n(\phi_{1}-\phi_{2}-\phi_{3}+\phi_{4})
\end{equation}
where $n$ is the order of the Bragg transitions involved and
$\alpha$ is the frequency chirping on the Bragg beams, used to
compensate the Doppler shift seen by the atoms during the
free-fall (typically set to a value of $\alpha = 42.5509$~kHz/ms,
for local gravity), while $\phi_{i}$ denotes the relative optical
phase between $i$-th Bragg pulse.

The trapped lattice phase will also induce an additional phase
shift $\phi_t=2k_\mathrm{blue} T_R N \times 2v_{r,\mathrm{green}}$, 
with respect to a free-falling atom \cite{cadoret2009,Charriere2012}, 
where $v_{r,\mathrm{green}}=\hbar k_\mathrm{green}/m$ is the recoil velocity 
of the green lattice. However, in order to compensate for the Doppler shift after 
the trapped atoms are released from the lattice, the Bragg laser needs to jump a
frequency $\Delta\omega = 2 k_\mathrm{blue} N \times 2v_{r,\mathrm{green}}$ 
rather than keeping a constant chirping rate $\alpha$ for a free-falling frame, 
which means the phase shift term $\phi_t$ can be exactly cancelled. 
A demonstration of the interferometer for first order Bragg pulses ($n=1$) is shown as an example in
Fig.~\ref{fig.fringes}, where the signal is showing the expected sinusoidal behaviour 
as a function of $\alpha$:
\begin{equation}
P=P_{0} + \frac{C}{2} \cos[(2k_\mathrm{blue} g-\alpha)T_{R}(T_{R}+T')].
\end{equation}

\begin{figure}\begin{center}
\includegraphics[width=0.47 \textwidth]{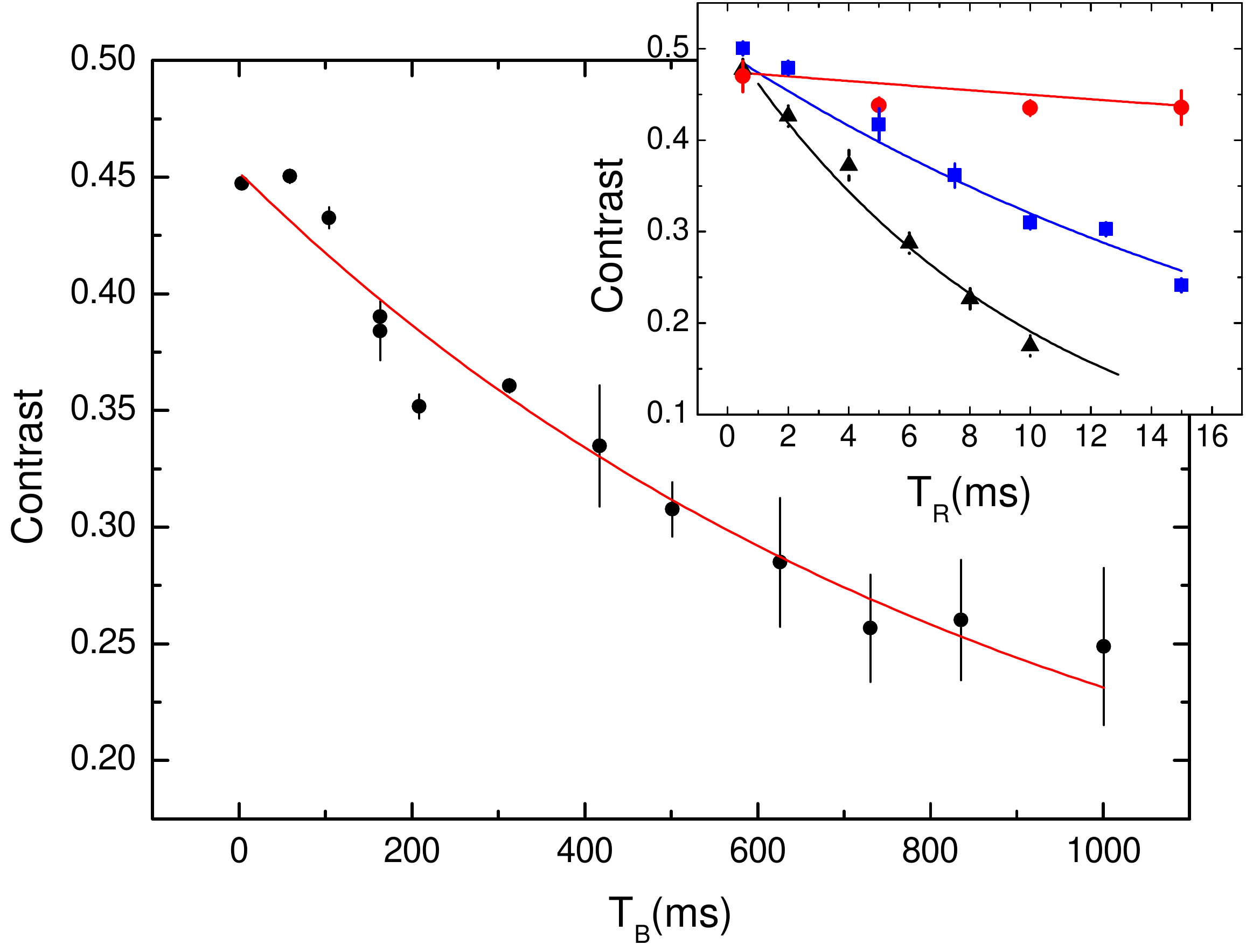}
\caption{Contrast decay as a function of Bloch evolution time
$T_{B}$. The separation time is $T_{R} = 1$~ms, and $T' = 37.8$~ms $+ T_{B}$.
The solid line is an exponential decay fit, giving
a time constant of 900(10)~ms. With this parameter choice the
interference is preserved for more than $N=575$ Bloch
oscillations. The inset shows the contrast as a function of $T_{R}$, 
with different sets of $T_{B}$ (red circles for $T_{B}$=3.5 ms, blue squares
for $T_{B}$=20.4 ms, and black triangles for $T_{B}$=59.1 ms). 
The solid lines in corresponding colours are exponential decay functions 
numerically estimated with the model of Eq.~\ref{eq.Random_kick}.   
} \label{fig.Contrast}
\end{center}
\end{figure}

The velocity splitting by a first order Bragg $\pi/2$ pulse is 17 mm/s, 
which for an interferometer with $T_{R} = 1$~ms results in a wave packet
separation of 0.017 mm. For the same Ramsey time $T_{R} =1$~ms, we also measured
the contrast evolution as a function of total Bloch oscillation time
$T_{B}$ (see Fig.~\ref{fig.Contrast}). For this measurement
the interval time between each set of $\pi/2$ pulse pairs is $T' =
37.8$~ms $+T_{B}$. The contrast decay with $T_{B}$ is satisfactorily fitted 
with an exponential decay function with a time constant of
$\tau=900(10)$ ms. In this condition, we show that the
interference can be preserved for a total number $N=575$ of Bloch
oscillations in the green lattice.

The fringe contrast of this lattice-trapped interferometer is an
important feature because it limits the potential performances for
future precision measurements of gravity and gravity gradients.
For this reason, we also studied the contrast decay as a function
of both parameters $T_R$ and $T_B$. In a second set of
measurements we fix the Bloch evolution time $T_{B}$, and we
recorded the contrast as a function of $\pi/2$ pulse separation
time $T_{R}$, as shown in the inset of Fig.~\ref{fig.Contrast}.

Following the argument in \cite{Charriere2012}, if we associate
this decay to a random variation of the longitudinal velocity $\delta v$ of the
atomic wave packet during the interferometer, for a
certain $T_{R}$ this induces a random phase shift of
$\delta\phi=2 k_\mathrm{blue} \delta v T_R$. Eventually, the contrast decay is
given by the convolution of this random phase shift $\delta\phi$
with the probability distribution of velocity variation $P(\delta
v)$:
\begin{equation}\label{eq.Random_kick}
\frac{C}{C_0}=\int P(\delta v)\cos(2nk_\mathrm{blue} T_R \delta v)\mathrm{d} \delta v
\end{equation}
where $n$ is the order of the Bragg transition. Applying this
formula to the contrast observed as a function of $T_{R}$ in the inset of
Fig.~\ref{fig.Contrast}, for $T_B$ = 3.5~ms, 20.4~ms, 59.1~ms, 
the distribution function $P(\delta v)$ is
nearly Lorentzian with a width $\gamma \sim $~0.2~$\mu$m/s, 1.6~$\mu$m/s, 
and 3.6~$\mu$m/s respectively, corresponding to very small momentum changes, 
less than $0.0004~ \hbar k_\mathrm{blue}$, depending on $T_B$.

\section{Decoherence sources}

In this section we present a detailed analysis of different decoherence
sources in our lattice-trapped interferometer. 
Most of the measurements focus on a precise determination of the atomic
momentum distribution, a key property strongly affecting the final
interferometer contrast \cite{Szigeti2012}. It is important to
notice that, as shown by the model in Eq.~\ref{eq.Random_kick},
it's not only the initial momentum spread of the atoms which contributes to
determining the maximum contrast achievable, 
but the momentum variation during the interferometer which
can also crucially affect the final interferometer outcome.
Since the system is not vibration isolated, a phase noise is present
on the Bragg and lattice laser beams. 
However at the resolution of our measurements, this phase noise does not affect 
the momentum distribution nor the coherence length in the Bloch evolution stage, 
as we will show in this section.

To independently study the various decoherence sources taking
place along the whole interferometer sequence, simpler
interferometer schemes were studied, such as pure Ramsey-Bord\'e 
or Mach-Zehnder configurations, as well as
measurements of the atomic momentum distribution during the
various phases of the full trapped interferometer sequence.

\subsection{Photon scattering}

A first source of decoherence resulting in contrast losses comes
from photon scattering events from off resonant blue and
green light involved in the interferometer.
Due to the large detuning of the 532 nm lattice light, we expect a
negligible contribution to the interferometer contrast decay.
Indeed, for a total light intensity of about 90~W/cm$^2$ at 532~nm we
estimate a photon scattering rate of $\Gamma_{s,\mathrm{green}}=0.007$
s$^{-1}$, allowing a long lattice trap lifetime and a long
coherence time for the Bloch oscillation phase.

A more important contribution comes from the Bragg laser light at 461~nm.
The Bragg laser is a semiconductor infrared source, which is amplified
with a tapered amplifier and then frequency doubled into the blue part of the optical spectrum. Although the laser is frequency stabilized
with a detuning of 10 GHz from $^{1}S_{0}$--$^{1}P_{1}$
transition at 461 nm, the optical output spectrum of the tapered amplifier
is several nm wide and the non-linear crystal employed for
frequency doubling could also double some frequency components close to
resonance. This could potentially cause additional amplified spontaneous emission
(ASE) at a rate $\Gamma_{ASE}$ \cite{andia2015}.

\begin{figure}\begin{center}
\includegraphics[width=0.47 \textwidth]{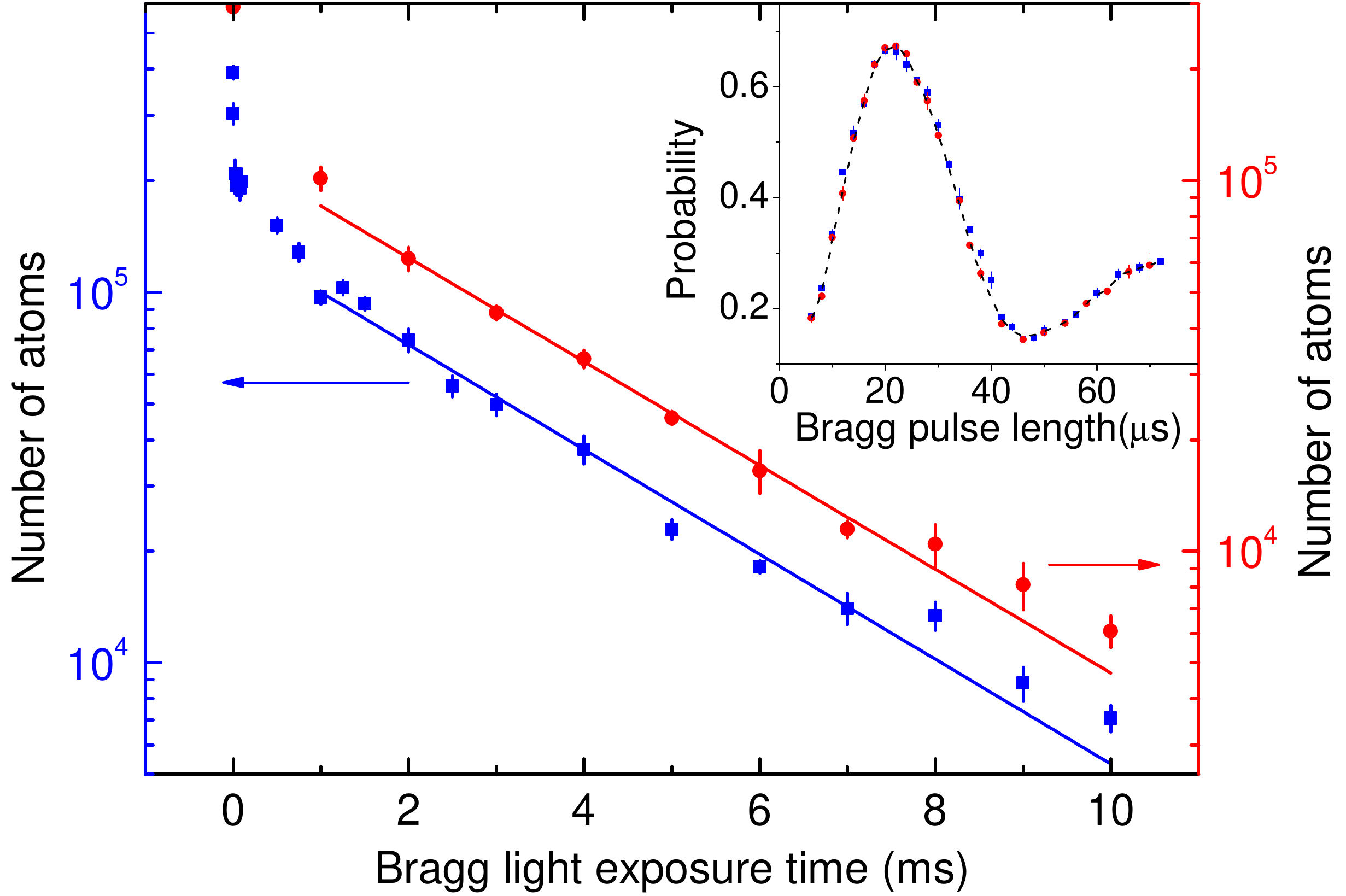}
\caption{Lattice decay measurements in the presence of Bragg light
	square pulse. The blue squares correspond to 150 \degree C
	heat-pipe temperature, and red circles to 400 \degree C. The upper blue and
	lower red solid lines are exponential decay fits for each respective set of data.
	The decay time constants according to the fits are 3.07 (13) ms 
	for 150 \degree C and 3.10 (13) ms for 400 \degree C.
	The two vertical axes are offset with respect to each other to
	display these two sets of data clearly. The insert plot is Rabi
	oscillations taken at 150\degree C and 400\degree C with the same
	plot notation. } \label{fig.SpEm}
\end{center}
\end{figure}

In order to avoid this additional resonance scattering, we
filtered the laser light from the Bragg source with a
30 cm-long strontium heat-pipe operating at 400\degree C. At this
temperature on-resonance light is absorbed by about 95\%
($-20$ dB). Several tests were done (as shown in
Fig.~\ref{fig.SpEm}) to study the effect of spontaneous emission
in our system, by keeping the Bragg laser 10 GHz detuned from
the atomic transition with an intensity of 400 mW/cm$^2$. Under these
conditions, the estimated scattering rate $\Gamma_{s,\mathrm{blue}}$ for a pure
monocromatic source, is $\Gamma_{s,\mathrm{blue}}=300$ Hz.

In a first test, we observed Rabi oscillations for an $n=2$ Bragg
transition for different heat-pipe temperatures, ranging from
$150\degree$C to $400\degree$C. As shown in the inset on
Fig.~\ref{fig.SpEm} the Rabi oscillations taken at different
heat-pipe temperatures are measured to be the same, from which we infer that
during a Bragg pulse $\Gamma_{ASE}$ is negligible.

To confirm this, a second set of measurements of the
lattice lifetime in presence of the Bragg beams, as a function of
heat-pipe temperature, were taken. As shown by the data in
Fig.~\ref{fig.SpEm}, the lattice decay time constants do not
depend on heat-pipe temperature, and the mean value
$\tau_s=3.08(13)$ ms is consistent with the previously estimated
resonant scattering rate $\Gamma_{s,\mathrm{blue}}$. 
This result is also confirmed by a third test, in which 
contrast measurements of a Mach-Zehnder interferometer have been 
repeated for the same temperature set. Also in this case the contrast 
observed is $C=70$\%, independent from heat-pipe temperature. 
Here, the contrast is mainly limited by Rabi frequency inhomogeneities
that arise from a combined effect of atomic cloud expansion and
imperfections on the profile of the Bragg beams
\cite{Muller2008,McDonald2013}. However, all additional
decoherence processes during the trapped Bloch evolution phase
will further decrease the value observed here in simple
Ramsey-Bord\'e or Mach-Zehnder configurations.

\subsection{Momentum distribution and lattice dynamics}

Further characterization of the atomic wave packet evolution
during the interferometer has been done in order to evidence small
random velocity changes, predicted to be responsible for the loss of
contrast.

In a first set of measurements we evaluate the coherence length of
the atomic sample for different interferometer sequences. This is
done by measuring the interferometer contrast as a
function of the time delay $\delta T$ on the last recombining
beam-splitter pulse. As the delay is increased, the fringe
contrast decays because of the reduced wave packet overlapping at
the recombination position. As a result, the typical contrast
envelope as a function of the time delay is described by a
Gaussian of the form \cite{Parazzoli2012}:
\begin{equation}\label{eq.Gauss_contrast}
C(\delta T) = C_0 + A \exp\left( {-\frac{v_r \delta T^2}{8
x_a^2}}\right)
\end{equation}
which represents the convolution of free-space Gaussian wave
packets with a coherence length $x_a$, recombining at position
$\delta x = v_r \delta T^2$, where $v_r = 2 \hbar n k_\mathrm{blue} /m$ is the
recoil velocity for an $n$-order Bragg transition.

\begin{figure}\begin{center}
\includegraphics[width=0.47 \textwidth]{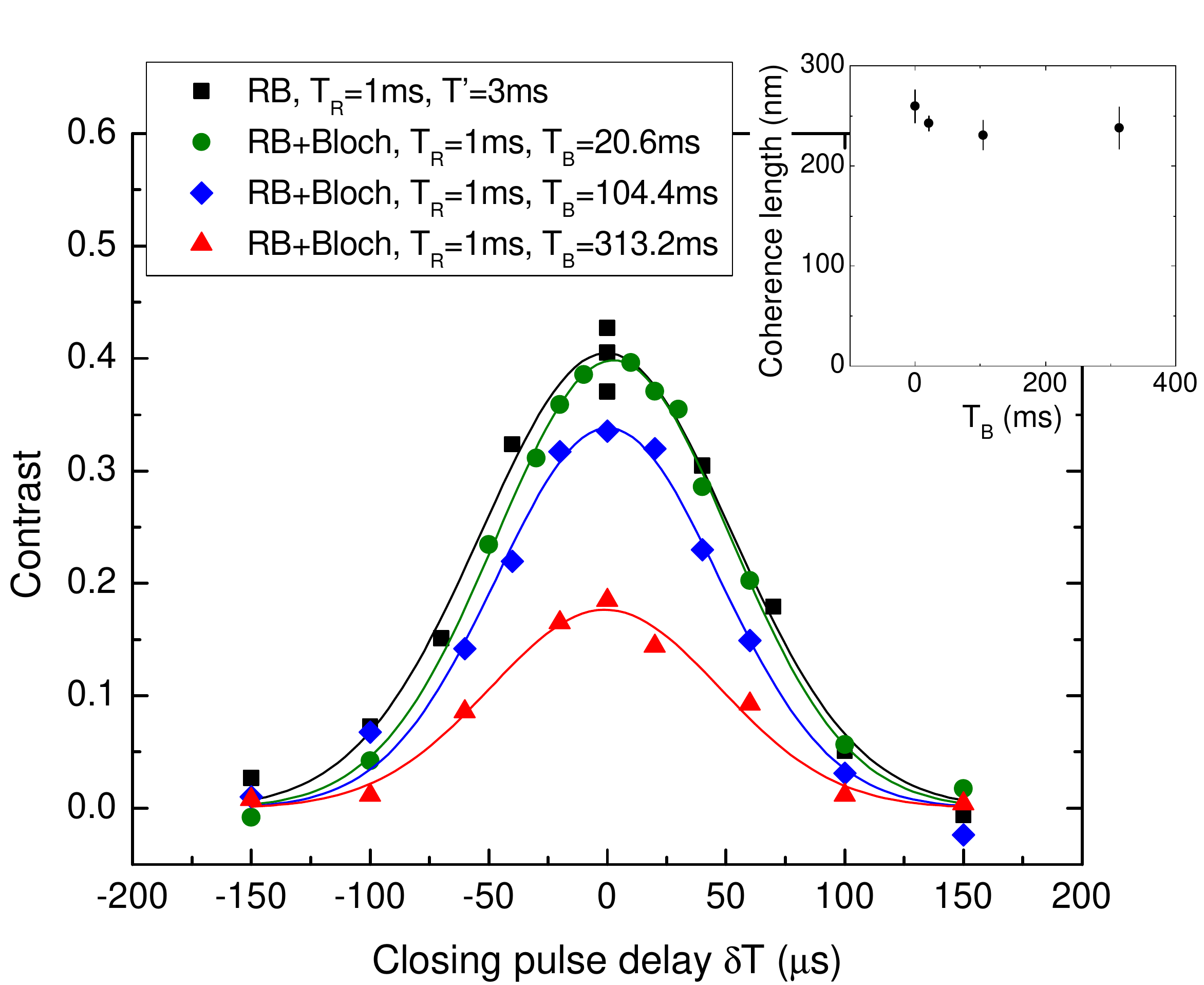}
\caption{Contrast measurements for different Bloch oscillation
periods $T_B$ (with Ramsey time $T_R = 1$ ms), as a function of
final pulse delay time $\delta T$. The contrast envelope for a
pure Ramsey-Bord\'e interferometer is also reported as reference.  
Offsets were removed for better comparison of the fitted curves.
The inset shows the estimated coherence length of the atomic
sample given by the Gaussian fit of the contrast
measurements.}\label{fig.Contrast_envelope}
\end{center}
\end{figure}

It is important to notice that in a free-space interferometer it
has been demonstrated that the coherence length is independent of the
wave packet's time evolution \cite{kellogg2007}. As a consequence,
the coherence length depends only on the initial longitudinal
velocity momentum distribution as given by the Heisenberg's uncertainty
principle. A similar approach has not yet been discussed for
trapped configurations, so we performed a set of measurement to
check whether the lattice might perturb the atomic coherence
length. Fig.~\ref{fig.Contrast_envelope} shows the results of the
contrast envelope measurements for different Bloch oscillation
durations $T_B$.  As a reference, the contrast envelope for a
pure Ramsey-Bord\'e interferometer is also reported.
The data are fitted with Gaussian functions as in
Eq.~\ref{eq.Gauss_contrast}. The fitted coherence lengths are
shown in the inset of Fig.~\ref{fig.Contrast_envelope} and the
results (up to $T_B=331.2$~ms) are consistent within 1~$\sigma$
with a mean value of $236(5)$~nm. Since the coherence length is
directly related to the velocity spread of the wave packets via
$\delta x \delta p =\hbar/2$, the resulting momentum spread is
$\delta p=0.155(4)$~$\hbar k_\mathrm{blue}$, a value consistent with
independent measurements done through Bragg spectroscopy (see the
next section). This analysis indicates that any changes in the
momentum distribution introduced by the trapped interferometer
phase is small and below the measurement sensitivity. In other
words, the Bloch oscillation phase does not increase the
longitudinal velocity distribution by an amount larger than 
previous estimation by the model in Eq.~\ref{eq.Random_kick} with
the observed contrast decay (see Fig.~\ref{fig.Contrast}).

\begin{figure}\begin{center}
\includegraphics[width=0.47 \textwidth]{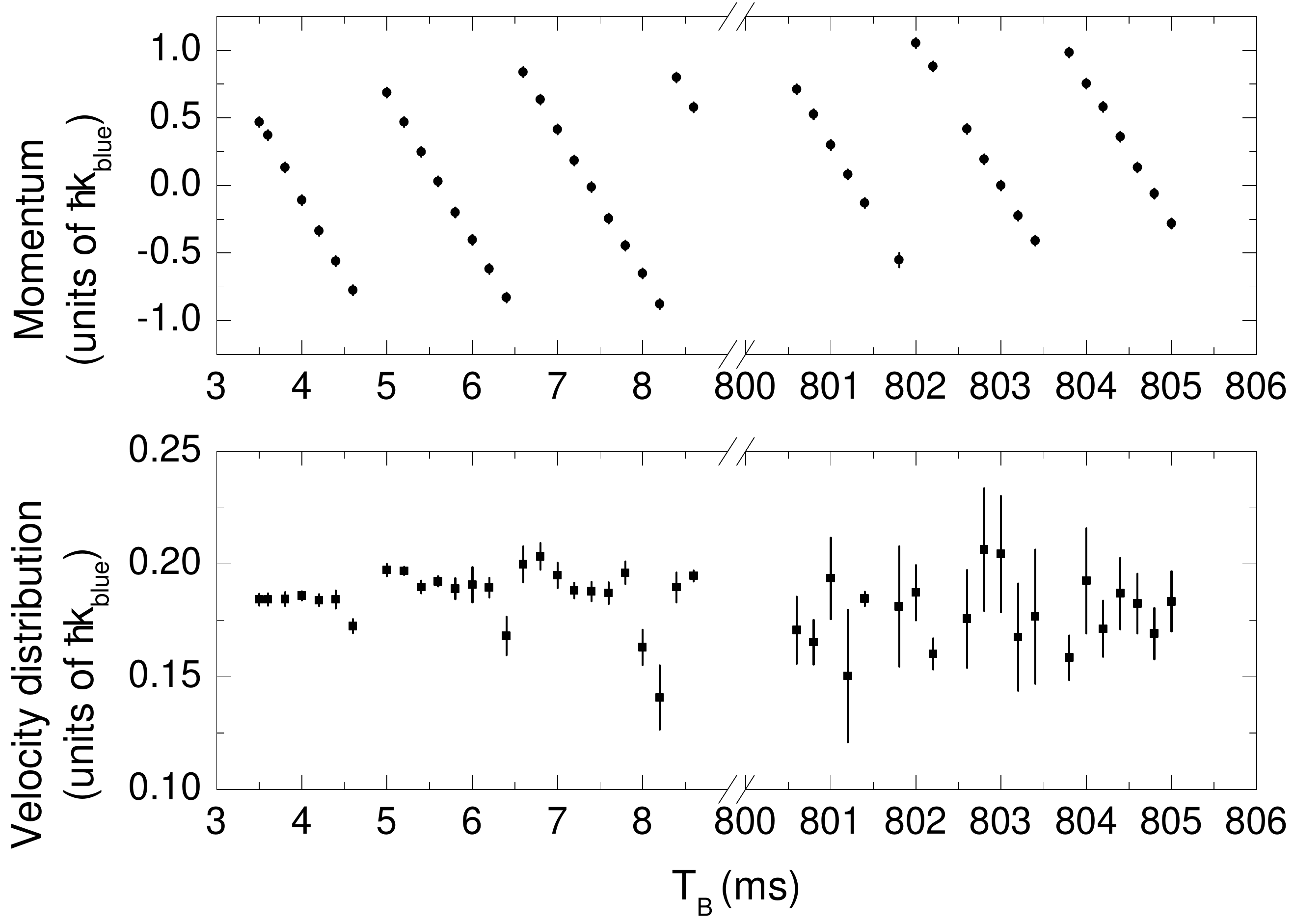}
\caption{Bloch oscillations in the recaptured lattice. The top
	plot is the calibrated TOF signal of $\sim$ 5 ms and $\sim$ 800 ms
	evolution time. Correspondingly, the bottom plot is the width of the
	velocity distribution of the atomic cloud.} 
\label{fig.RecapBloch}
\end{center}
\end{figure}

A similar result can be obtained independently from the direct
measurement of the atomic momentum distribution during the various
interferometer phases, and in particular after the Bloch
oscillation phase. In this case, Bloch oscillations have been
observed in TOF, and information on the atomic
cloud's longitudinal velocity distribution has been extracted.
Fig.~\ref{fig.RecapBloch} shows the Bloch oscillations in the
recaptured lattice up to 800 ms evolution time, together (bottom
plot) with the estimated width of the atomic longitudinal velocity
distribution. The data shows that during the Bloch evolution, the
mean longitudinal velocity width does not change. The large
standard deviations on distribution widths for longer evolution
times arise mainly from fitting errors, which increase for
smaller atomic number.

\begin{figure}\begin{center}
\includegraphics[width=0.47 \textwidth]{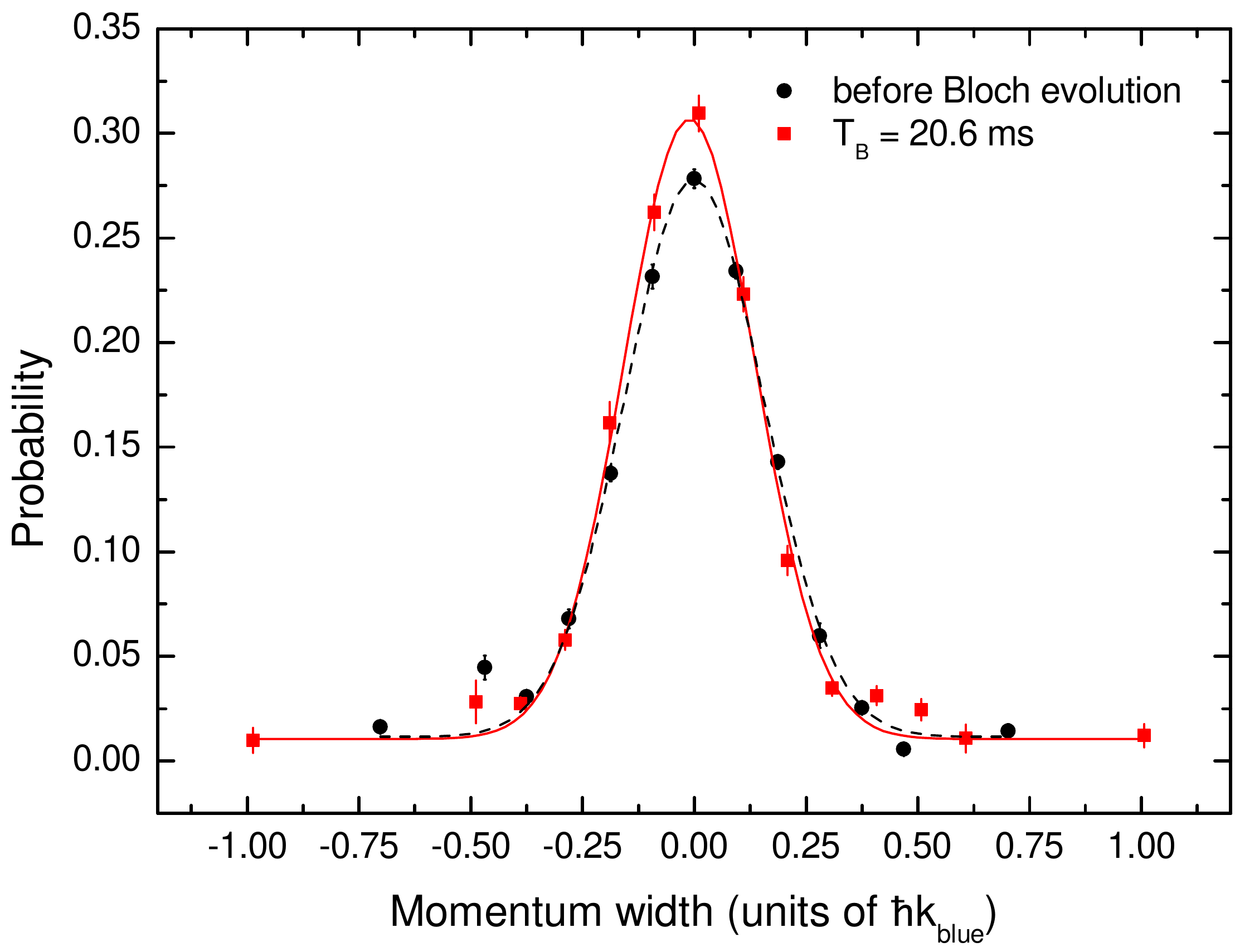}
\caption{Bragg spectroscopy of the atomic cloud before Bloch evolution
	(black circles), and after 20.6 ms Bloch evolution (red squares).
	The dashed black line and solid red line are  Gaussian amplitude fits of the respective
	data set. } \label{fig.Braggspec}
\end{center}
\end{figure}

An even more sensitive technique employed for the estimation of the
atomic cloud velocity distribution is Bragg spectroscopy. Here,
Bragg spectra were recorded before the lattice recapture and
after 20.6 ms Bloch evolution, as shown in
Fig.~\ref{fig.Braggspec}. The Bragg pulse for spectroscopy has a
Gaussian time profile with a duration of 150 $\mu$s. The momentum
resolution of this Bragg pulse is 0.02~$\hbar k_\mathrm{blue}$.
The fit of the two datasets gives a momentum width of
0.155(4)~$\hbar k_\mathrm{blue}$ and 0.147(4)~$\hbar k_\mathrm{blue}$,
before and after the Bloch oscillation phase respectively. Again,
the difference between the two values is within the resolution of
the Bragg spectroscopy itself, indicating that there is no
process changing the momentum distribution
during the Bloch evolution phase. The measured velocity
distribution by TOF signal of Bloch oscillations is consistent
with the one obtained by Bragg spectroscopy.

Again, the study of the lattice dynamics in terms of momentum
spread confirms the same result: the interferometer contrast decay
is mainly due to very small random changes in the atomic velocity
that are well below the measurement resolution in all the
experimental tests mentioned above.

\subsection{Random momentum changes from lattice speckles}

Despite being below our present level of measurement resolution, as
shown in the previous section, we searched for possible cause
of the small momentum changes in our interferometer other than
photon scattering events. Especially during the long
lattice phase, random momentum changes may arise from random
lattice intensity gradients which, through dipole forces on
the atoms, can impart a momentum perturbation to the
interferometer and eventually cause a contrast loss.

\begin{figure}\begin{center}
\includegraphics[width=0.47 \textwidth]{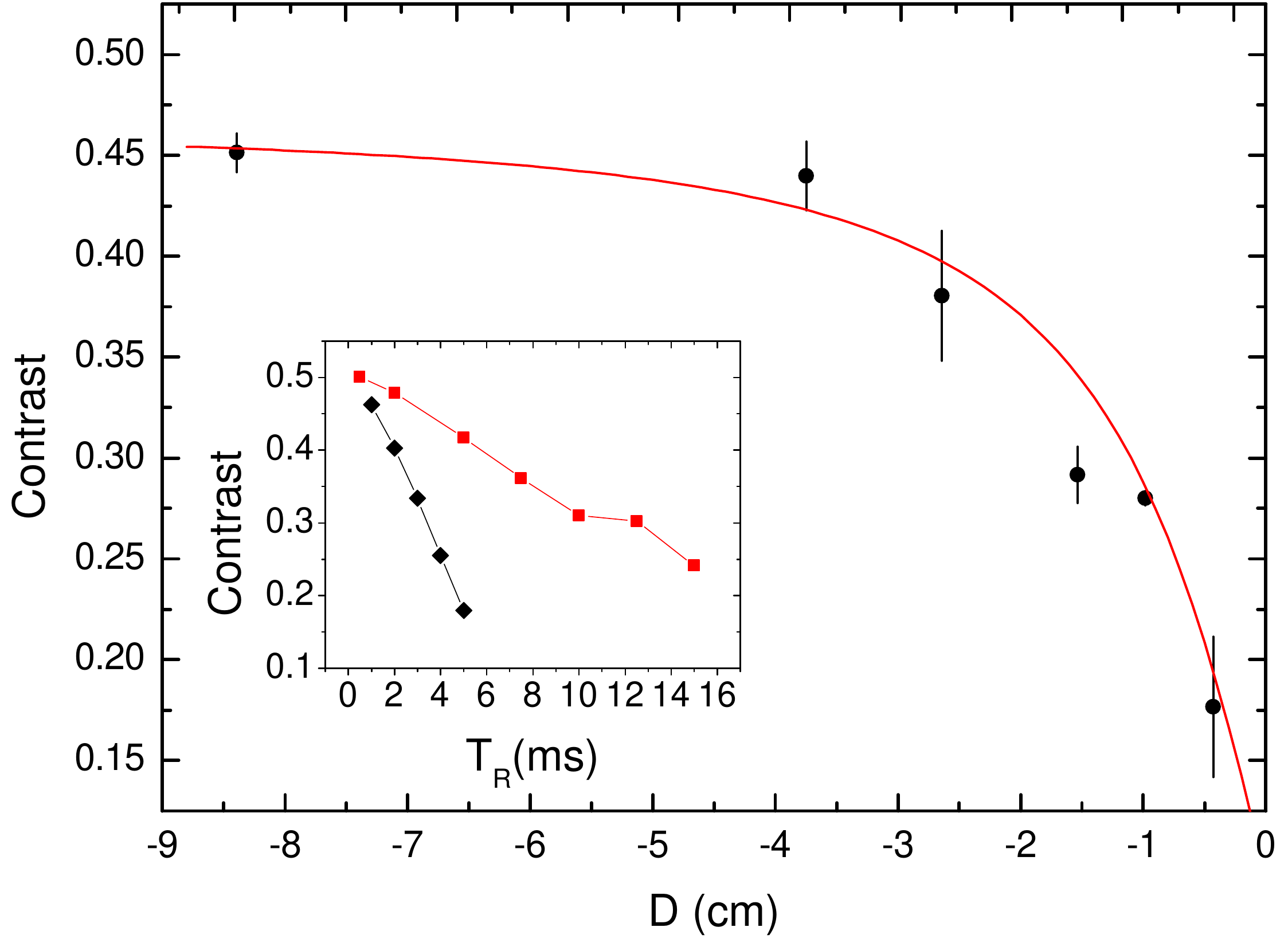}
\caption{The black circles are the contrast loss measurements
	versus the distance $D$ between the atomic cloud and the top
	window of the vacuum chamber (the minus sign means the atoms are 
	below the window), the solid line is a numerical estimation with
    Eq.~\ref{eq.Random_kick}, using a Lorentzian distribution with a
	width scaling as $D^{-2}$. For this measurement we choose $T_R =1$~ms 
	and $T_B = 38.4$~ms. The inset plot is the contrast decay as
	a function of $T_R$, for a fixed $T_B = 20.4$~ms, for two
	different lattice beam waists. Black diamonds correspond to beam waist 
	of $300\,\mu$m, and red squares with $800\,\mu$m 
	(The solid lines are for eye-guide).}
\label{fig.Contrast_speckle}
\end{center}
\end{figure}

In particular, a speckle pattern arising from imperfections in some
optical elements (especially the vacuum chamber windows) is 
the largest contributor to this effect. To show clearly how this
effect can dramatically affect the interferometer, we performed a
measurement of the contrast loss as a function of the distance $D$
between the trapped atoms and the top window of the vacuum
chamber (the lattice beam waist was $\sim 300~\mu$m for this measurement). 
In this case, the trapping position has been
controlled by modifying the lattice elevation time. By
classical geometrical consideration, the dipole force induced by
speckles produced by the window scales as $D^{-2}$, therefore,
given a fixed Bloch evolution time, we expect a similar dependence
of the random velocity variation as a function of the distance
$D$. The result of the measurement is shown in
Fig.~\ref{fig.Contrast_speckle}. Here we numerically estimated the contrast
decay with Eq.~\ref{eq.Random_kick}, with random velocity variation
$\delta v= (\delta U_0 d^2 /16\lambda D^2)(T_B / m) $ \cite{charriere2011}, 
where $\delta U_0$ is the potential depth variation 
due to speckle pattern and $d$ the dimension of the
imperfection on the window. For our system $\delta U_0 \sim$ 0.05 $U_0$, $d \sim
250~\mu$m. The result of this model explains the contrast decay in
Fig.~\ref{fig.Contrast_speckle} and gives
an estimation of $\delta v \sim 1.2~ \mu$m/s for the atoms at the 
position  of our typical experimental sequence (with $D = 8.4$ cm and $T_B = 38.4$ ms), consistent with the
estimation obtained from the contrast decay as a function of $T_R$ in section III.

While in a typical experimental sequence the atoms are held
sufficiently far from the top and bottom windows,
the previous measurement shows clearly that any small intensity
imperfection in the lattice profile can dramatically reduce the
interferometer contrast. Far from the windows, the Gaussian beam profile
depends strongly on the optical setup chosen to collimate the beam
as well as on diffraction from apertures along the beam path.
Furthermore, given a finite Rayleigh length (proportional to the square of the
beam waist) of the lattice beams, it is also possible to have a coupling between
the transverse and longitudinal motions in the lattice due to the wavefront
curvature. One clear indication of this fact is shown by a measurement
of the contrast decay as a function of the lattice beam waist.
For this measurement we choose $T_B = 20.4$~ms and we measured
the contrast decay as a function of the Ramsey time $T_R$ for two
different lattice beam waists, 300 $\mu$m and
$800\,\mu$m respectively. As shown in the inset of
Fig~\ref{fig.Contrast_speckle}, the contrast is improved by a
factor of about $4$ (for a fixed value of $T_R = 5$ ms), when the lattice
beam waist at the atoms' position is enlarged by a factor of $\sim 2$.

The contrast decay is therefore strongly affected by the lattice beam
profile and it can be improved dramatically by increasing the
lattice size. From this evidence, we expect that a further
increase of the lattice beam size or a better optical setup may
improve our system. In the present experimental configuration this
is not possible due to the limited lattice laser power
at 532~nm. Another possible approach is using a resonant
cavity for both the lattice and interferometer beams, 
so that the beam profile can be improved, and 
moreover high laser intensity can be achieved \cite{Hamilton2015}.

\section{Conclusions}

In conclusion, we demonstrated a high contrast atom interferometer
based on a combination of Ramsey-Bord\'e Bragg pulses and a Bloch
oscillation stage in a vertical lattice. By using $^{88}$Sr atoms,
the total interferometer time we reach is 1 s, currently limited
by the atom lifetime in the lattice and geometry imperfections of the lattice
beams.
This limitation has been highlighted through a detailed
experimental study of main decoherence sources in our system. In
particular, a detailed study of the contrast decay sources as
functions of the principal parameters (as the pulse separation time $T_R$
and the Bloch evolution time $T_B$) has been conducted.

As a result, we evidenced that at this level, decoherence is
mainly given by technical and not fundamental limitations. For this
reason, we think that trapped interferometer schemes employing
Bloch oscillations with $^{88}$Sr atoms are valid candidates for
further extending the total atom interferometer time, towards high
precision inertial measurements.

A future prospect is to use a red
laser at 689 nm to both drive the Bragg transitions and to trap
the atoms in a standing wave lattice. Thanks to the narrow
intercombination line and the high power available at this wavelength,
it should be possible to realize larger trapping beams maintaining a
sufficiently low photon scattering rate.

\subsection*{Acknowledgments} 
We thank Leonardo Salvi for help in the early stage of the experiment.
We acknowledge financial support
from INFN and the Italian Ministry of Education, University
and Research (MIUR) under the Progetto Premiale ``Interferometro
Atomico''. We also acknowledge support from the European Union's
Seventh Framework Programme (FP7/2007-2013 grant agreement 250072
- ``iSense'' project and grant agreement 607493 - ITN ``FACT''
project).


\begin{thebibliography}{35}%
\makeatletter
\providecommand \@ifxundefined [1]{%
 \@ifx{#1\undefined}
}%
\providecommand \@ifnum [1]{%
 \ifnum #1\expandafter \@firstoftwo
 \else \expandafter \@secondoftwo
 \fi
}%
\providecommand \@ifx [1]{%
 \ifx #1\expandafter \@firstoftwo
 \else \expandafter \@secondoftwo
 \fi
}%
\providecommand \natexlab [1]{#1}%
\providecommand \enquote  [1]{``#1''}%
\providecommand \bibnamefont  [1]{#1}%
\providecommand \bibfnamefont [1]{#1}%
\providecommand \citenamefont [1]{#1}%
\providecommand \href@noop [0]{\@secondoftwo}%
\providecommand \href [0]{\begingroup \@sanitize@url \@href}%
\providecommand \@href[1]{\@@startlink{#1}\@@href}%
\providecommand \@@href[1]{\endgroup#1\@@endlink}%
\providecommand \@sanitize@url [0]{\catcode `\\12\catcode `\$12\catcode
  `\&12\catcode `\#12\catcode `\^12\catcode `\_12\catcode `\%12\relax}%
\providecommand \@@startlink[1]{}%
\providecommand \@@endlink[0]{}%
\providecommand \url  [0]{\begingroup\@sanitize@url \@url }%
\providecommand \@url [1]{\endgroup\@href {#1}{\urlprefix }}%
\providecommand \urlprefix  [0]{URL }%
\providecommand \Eprint [0]{\href }%
\providecommand \doibase [0]{http://dx.doi.org/}%
\providecommand \selectlanguage [0]{\@gobble}%
\providecommand \bibinfo  [0]{\@secondoftwo}%
\providecommand \bibfield  [0]{\@secondoftwo}%
\providecommand \translation [1]{[#1]}%
\providecommand \BibitemOpen [0]{}%
\providecommand \bibitemStop [0]{}%
\providecommand \bibitemNoStop [0]{.\EOS\space}%
\providecommand \EOS [0]{\spacefactor3000\relax}%
\providecommand \BibitemShut  [1]{\csname bibitem#1\endcsname}%
\let\auto@bib@innerbib\@empty
\bibitem [{tin(2014)}]{tino2013atom}%
  \BibitemOpen
  \href@noop {} {\bibfield  {journal} {\bibinfo  {journal} {\emph{Atom
  interferometry}, edited by G. M. Tino and M. A. Kasevich}\ } (\bibinfo {year}
  {Societ\`a Italiana di Fisica and IOS Press, Amsterdam, 2014})}\BibitemShut
  {NoStop}%
\bibitem [{\citenamefont {Gillot}\ \emph {et~al.}(2013)\citenamefont {Gillot},
  \citenamefont {Lepoutre}, \citenamefont {Gauguet}, \citenamefont
  {B\"uchner},\ and\ \citenamefont {Vigu\'e}}]{Gillot2013}%
  \BibitemOpen
  \bibfield  {author} {\bibinfo {author} {\bibfnamefont {J.}~\bibnamefont
  {Gillot}}, \bibinfo {author} {\bibfnamefont {S.}~\bibnamefont {Lepoutre}},
  \bibinfo {author} {\bibfnamefont {A.}~\bibnamefont {Gauguet}}, \bibinfo
  {author} {\bibfnamefont {M.}~\bibnamefont {B\"uchner}}, \ and\ \bibinfo
  {author} {\bibfnamefont {J.}~\bibnamefont {Vigu\'e}},\ }\href {\doibase
  10.1103/PhysRevLett.111.030401} {\bibfield  {journal} {\bibinfo  {journal}
  {Phys. Rev. Lett.}\ }\textbf {\bibinfo {volume} {111}},\ \bibinfo {pages}
  {030401} (\bibinfo {year} {2013})}\BibitemShut {NoStop}%
\bibitem [{\citenamefont {Kovachy}\ \emph {et~al.}(2015)\citenamefont
  {Kovachy}, \citenamefont {Asenbaum}, \citenamefont {Overstreet},
  \citenamefont {Donnelly}, \citenamefont {Dickerson}, \citenamefont
  {Sugarbaker}, \citenamefont {Hogan},\ and\ \citenamefont
  {Kasevich}}]{Kovachy2015}%
  \BibitemOpen
  \bibfield  {author} {\bibinfo {author} {\bibfnamefont {T.}~\bibnamefont
  {Kovachy}}, \bibinfo {author} {\bibfnamefont {P.}~\bibnamefont {Asenbaum}},
  \bibinfo {author} {\bibfnamefont {C.}~\bibnamefont {Overstreet}}, \bibinfo
  {author} {\bibfnamefont {C.~A.}\ \bibnamefont {Donnelly}}, \bibinfo {author}
  {\bibfnamefont {S.~M.}\ \bibnamefont {Dickerson}}, \bibinfo {author}
  {\bibfnamefont {A.}~\bibnamefont {Sugarbaker}}, \bibinfo {author}
  {\bibfnamefont {J.~M.}\ \bibnamefont {Hogan}}, \ and\ \bibinfo {author}
  {\bibfnamefont {M.~A.}\ \bibnamefont {Kasevich}},\ }\href {\doibase
  {10.1038/nature16155}} {\bibfield  {journal} {\bibinfo  {journal} {{Nature
  (London)}}\ }\textbf {\bibinfo {volume} {{528}}},\ \bibinfo {pages} {{530}}
  (\bibinfo {year} {{2015}})}\BibitemShut {NoStop}%
\bibitem [{\citenamefont {Lopes}\ \emph {et~al.}(2015)\citenamefont {Lopes},
  \citenamefont {Imanaliev}, \citenamefont {Aspect}, \citenamefont {Cheneau},
  \citenamefont {Boiron},\ and\ \citenamefont {Westbrook}}]{Lopes2015}%
  \BibitemOpen
  \bibfield  {author} {\bibinfo {author} {\bibfnamefont {R.}~\bibnamefont
  {Lopes}}, \bibinfo {author} {\bibfnamefont {A.}~\bibnamefont {Imanaliev}},
  \bibinfo {author} {\bibfnamefont {A.}~\bibnamefont {Aspect}}, \bibinfo
  {author} {\bibfnamefont {M.}~\bibnamefont {Cheneau}}, \bibinfo {author}
  {\bibfnamefont {D.}~\bibnamefont {Boiron}}, \ and\ \bibinfo {author}
  {\bibfnamefont {C.~I.}\ \bibnamefont {Westbrook}},\ }\href@noop {} {\bibfield
   {journal} {\bibinfo  {journal} {Nature (London)}\ }\textbf {\bibinfo
  {volume} {520}},\ \bibinfo {pages} {66} (\bibinfo {year} {2015})}\BibitemShut
  {NoStop}%
\bibitem [{\citenamefont {Manning}\ \emph {et~al.}(2015)\citenamefont
  {Manning}, \citenamefont {Khakimov}, \citenamefont {Dall},\ and\
  \citenamefont {Truscott}}]{Manning2015}%
  \BibitemOpen
  \bibfield  {author} {\bibinfo {author} {\bibfnamefont {A.}~\bibnamefont
  {Manning}}, \bibinfo {author} {\bibfnamefont {R.}~\bibnamefont {Khakimov}},
  \bibinfo {author} {\bibfnamefont {R.}~\bibnamefont {Dall}}, \ and\ \bibinfo
  {author} {\bibfnamefont {A.}~\bibnamefont {Truscott}},\ }\href@noop {}
  {\bibfield  {journal} {\bibinfo  {journal} {Nat. Phys}\ }\textbf {\bibinfo
  {volume} {11}},\ \bibinfo {pages} {539} (\bibinfo {year} {2015})}\BibitemShut
  {NoStop}%
\bibitem [{\citenamefont {Amelino-Camelia}\ \emph {et~al.}(2009)\citenamefont
  {Amelino-Camelia}, \citenamefont {L\"ammerzahl}, \citenamefont {Mercati},\
  and\ \citenamefont {Tino}}]{Amelino-Camelia2009}%
  \BibitemOpen
  \bibfield  {author} {\bibinfo {author} {\bibfnamefont {G.}~\bibnamefont
  {Amelino-Camelia}}, \bibinfo {author} {\bibfnamefont {C.}~\bibnamefont
  {L\"ammerzahl}}, \bibinfo {author} {\bibfnamefont {F.}~\bibnamefont
  {Mercati}}, \ and\ \bibinfo {author} {\bibfnamefont {G.~M.}\ \bibnamefont
  {Tino}},\ }\href {\doibase 10.1103/PhysRevLett.103.171302} {\bibfield
  {journal} {\bibinfo  {journal} {Phys. Rev. Lett.}\ }\textbf {\bibinfo
  {volume} {103}},\ \bibinfo {pages} {171302} (\bibinfo {year}
  {2009})}\BibitemShut {NoStop}%
\bibitem [{\citenamefont {Biedermann}\ \emph {et~al.}(2015)\citenamefont
  {Biedermann}, \citenamefont {Wu}, \citenamefont {Deslauriers}, \citenamefont
  {Roy}, \citenamefont {Mahadeswaraswamy},\ and\ \citenamefont
  {Kasevich}}]{Biedermann2015}%
  \BibitemOpen
  \bibfield  {author} {\bibinfo {author} {\bibfnamefont {G.~W.}\ \bibnamefont
  {Biedermann}}, \bibinfo {author} {\bibfnamefont {X.}~\bibnamefont {Wu}},
  \bibinfo {author} {\bibfnamefont {L.}~\bibnamefont {Deslauriers}}, \bibinfo
  {author} {\bibfnamefont {S.}~\bibnamefont {Roy}}, \bibinfo {author}
  {\bibfnamefont {C.}~\bibnamefont {Mahadeswaraswamy}}, \ and\ \bibinfo
  {author} {\bibfnamefont {M.~A.}\ \bibnamefont {Kasevich}},\ }\href {\doibase
  10.1103/PhysRevA.91.033629} {\bibfield  {journal} {\bibinfo  {journal} {Phys.
  Rev. A}\ }\textbf {\bibinfo {volume} {91}},\ \bibinfo {pages} {033629}
  (\bibinfo {year} {2015})}\BibitemShut {NoStop}%
\bibitem [{\citenamefont {Hartwig}\ \emph {et~al.}(2015)\citenamefont
  {Hartwig}, \citenamefont {Abend}, \citenamefont {Schubert}, \citenamefont
  {Schlippert}, \citenamefont {Ahlers}, \citenamefont {Posso-Trujillo},
  \citenamefont {Gaaloul}, \citenamefont {Ertmer},\ and\ \citenamefont
  {Rasel}}]{Hartwig2015}%
  \BibitemOpen
  \bibfield  {author} {\bibinfo {author} {\bibfnamefont {J.}~\bibnamefont
  {Hartwig}}, \bibinfo {author} {\bibfnamefont {S.}~\bibnamefont {Abend}},
  \bibinfo {author} {\bibfnamefont {C.}~\bibnamefont {Schubert}}, \bibinfo
  {author} {\bibfnamefont {D.}~\bibnamefont {Schlippert}}, \bibinfo {author}
  {\bibfnamefont {H.}~\bibnamefont {Ahlers}}, \bibinfo {author} {\bibfnamefont
  {K.}~\bibnamefont {Posso-Trujillo}}, \bibinfo {author} {\bibfnamefont
  {N.}~\bibnamefont {Gaaloul}}, \bibinfo {author} {\bibfnamefont
  {W.}~\bibnamefont {Ertmer}}, \ and\ \bibinfo {author} {\bibfnamefont {E.~M.}\
  \bibnamefont {Rasel}},\ }\href
  {http://stacks.iop.org/1367-2630/17/i=3/a=035011} {\bibfield  {journal}
  {\bibinfo  {journal} {New J. Phys}\ }\textbf {\bibinfo {volume} {17}},\
  \bibinfo {pages} {035011} (\bibinfo {year} {2015})}\BibitemShut {NoStop}%
\bibitem [{\citenamefont {Zhou}\ \emph {et~al.}(2015)\citenamefont {Zhou},
  \citenamefont {Long}, \citenamefont {Tang}, \citenamefont {Chen},
  \citenamefont {Gao}, \citenamefont {Peng}, \citenamefont {Duan},
  \citenamefont {Zhong}, \citenamefont {Xiong}, \citenamefont {Wang},
  \citenamefont {Zhang},\ and\ \citenamefont {Zhan}}]{Zhou2015}%
  \BibitemOpen
  \bibfield  {author} {\bibinfo {author} {\bibfnamefont {L.}~\bibnamefont
  {Zhou}}, \bibinfo {author} {\bibfnamefont {S.-T.}\ \bibnamefont {Long}},
  \bibinfo {author} {\bibfnamefont {B.}~\bibnamefont {Tang}}, \bibinfo {author}
  {\bibfnamefont {X.}~\bibnamefont {Chen}}, \bibinfo {author} {\bibfnamefont
  {F.}~\bibnamefont {Gao}}, \bibinfo {author} {\bibfnamefont {W.-C.}\
  \bibnamefont {Peng}}, \bibinfo {author} {\bibfnamefont {W.-T.}\ \bibnamefont
  {Duan}}, \bibinfo {author} {\bibfnamefont {J.-Q.}\ \bibnamefont {Zhong}},
  \bibinfo {author} {\bibfnamefont {Z.-Y.}\ \bibnamefont {Xiong}}, \bibinfo
  {author} {\bibfnamefont {J.}~\bibnamefont {Wang}}, \bibinfo {author}
  {\bibfnamefont {Y.-Z.}\ \bibnamefont {Zhang}}, \ and\ \bibinfo {author}
  {\bibfnamefont {M.-S.}\ \bibnamefont {Zhan}},\ }\href {\doibase
  10.1103/PhysRevLett.115.013004} {\bibfield  {journal} {\bibinfo  {journal}
  {Phys. Rev. Lett.}\ }\textbf {\bibinfo {volume} {115}},\ \bibinfo {pages}
  {013004} (\bibinfo {year} {2015})}\BibitemShut {NoStop}%
\bibitem [{\citenamefont {Duan}\ \emph {et~al.}(2016)\citenamefont {Duan},
  \citenamefont {Deng}, \citenamefont {Zhou}, \citenamefont {Zhang},
  \citenamefont {Xu}, \citenamefont {Xiong}, \citenamefont {Xu}, \citenamefont
  {Shao}, \citenamefont {Luo},\ and\ \citenamefont {Hu}}]{Duan2016}%
  \BibitemOpen
  \bibfield  {author} {\bibinfo {author} {\bibfnamefont {X.-C.}\ \bibnamefont
  {Duan}}, \bibinfo {author} {\bibfnamefont {X.-B.}\ \bibnamefont {Deng}},
  \bibinfo {author} {\bibfnamefont {M.-K.}\ \bibnamefont {Zhou}}, \bibinfo
  {author} {\bibfnamefont {K.}~\bibnamefont {Zhang}}, \bibinfo {author}
  {\bibfnamefont {W.-J.}\ \bibnamefont {Xu}}, \bibinfo {author} {\bibfnamefont
  {F.}~\bibnamefont {Xiong}}, \bibinfo {author} {\bibfnamefont {Y.-Y.}\
  \bibnamefont {Xu}}, \bibinfo {author} {\bibfnamefont {C.-G.}\ \bibnamefont
  {Shao}}, \bibinfo {author} {\bibfnamefont {J.}~\bibnamefont {Luo}}, \ and\
  \bibinfo {author} {\bibfnamefont {Z.-K.}\ \bibnamefont {Hu}},\ }\href
  {\doibase 10.1103/PhysRevLett.117.023001} {\bibfield  {journal} {\bibinfo
  {journal} {Phys. Rev. Lett.}\ }\textbf {\bibinfo {volume} {117}},\ \bibinfo
  {pages} {023001} (\bibinfo {year} {2016})}\BibitemShut {NoStop}%
\bibitem [{\citenamefont {Bouchendira}\ \emph {et~al.}(2011)\citenamefont
  {Bouchendira}, \citenamefont {Clad\'e}, \citenamefont {Guellati-Kh\'elifa},
  \citenamefont {Nez},\ and\ \citenamefont {Biraben}}]{Bouchendira2011}%
  \BibitemOpen
  \bibfield  {author} {\bibinfo {author} {\bibfnamefont {R.}~\bibnamefont
  {Bouchendira}}, \bibinfo {author} {\bibfnamefont {P.}~\bibnamefont
  {Clad\'e}}, \bibinfo {author} {\bibfnamefont {S.}~\bibnamefont
  {Guellati-Kh\'elifa}}, \bibinfo {author} {\bibfnamefont {F.}~\bibnamefont
  {Nez}}, \ and\ \bibinfo {author} {\bibfnamefont {F.}~\bibnamefont
  {Biraben}},\ }\href {\doibase 10.1103/PhysRevLett.106.080801} {\bibfield
  {journal} {\bibinfo  {journal} {Phys. Rev. Lett.}\ }\textbf {\bibinfo
  {volume} {106}},\ \bibinfo {pages} {080801} (\bibinfo {year}
  {2011})}\BibitemShut {NoStop}%
\bibitem [{\citenamefont {Rosi}\ \emph {et~al.}(2014)\citenamefont {Rosi},
  \citenamefont {Sorrentino}, \citenamefont {Cacciapuoti}, \citenamefont
  {Prevedelli},\ and\ \citenamefont {Tino}}]{Rosi2014}%
  \BibitemOpen
  \bibfield  {author} {\bibinfo {author} {\bibfnamefont {G.}~\bibnamefont
  {Rosi}}, \bibinfo {author} {\bibfnamefont {F.}~\bibnamefont {Sorrentino}},
  \bibinfo {author} {\bibfnamefont {L.}~\bibnamefont {Cacciapuoti}}, \bibinfo
  {author} {\bibfnamefont {M.}~\bibnamefont {Prevedelli}}, \ and\ \bibinfo
  {author} {\bibfnamefont {G.~M.}\ \bibnamefont {Tino}},\ }\href {\doibase
  {10.1038/nature13433}} {\bibfield  {journal} {\bibinfo  {journal} {{Nature}}\
  }\textbf {\bibinfo {volume} {{510}}},\ \bibinfo {pages} {{518}} (\bibinfo
  {year} {{2014}})}\BibitemShut {NoStop}%
\bibitem [{\citenamefont {Tino}\ and\ \citenamefont
  {Vetrano}(2007)}]{tino2007p}%
  \BibitemOpen
  \bibfield  {author} {\bibinfo {author} {\bibfnamefont {G.~M.}\ \bibnamefont
  {Tino}}\ and\ \bibinfo {author} {\bibfnamefont {F.}~\bibnamefont {Vetrano}},\
  }\href@noop {} {\bibfield  {journal} {\bibinfo  {journal} {Class. Quantum
  Grav}\ }\textbf {\bibinfo {volume} {24}},\ \bibinfo {pages} {2167} (\bibinfo
  {year} {2007})}\BibitemShut {NoStop}%
\bibitem [{\citenamefont {Dimopoulos}\ \emph {et~al.}(2009)\citenamefont
  {Dimopoulos}, \citenamefont {Graham}, \citenamefont {Hogan}, \citenamefont
  {Kasevich},\ and\ \citenamefont {Rajendran}}]{Dimopoulos2009}%
  \BibitemOpen
  \bibfield  {author} {\bibinfo {author} {\bibfnamefont {S.}~\bibnamefont
  {Dimopoulos}}, \bibinfo {author} {\bibfnamefont {P.~W.}\ \bibnamefont
  {Graham}}, \bibinfo {author} {\bibfnamefont {J.~M.}\ \bibnamefont {Hogan}},
  \bibinfo {author} {\bibfnamefont {M.~A.}\ \bibnamefont {Kasevich}}, \ and\
  \bibinfo {author} {\bibfnamefont {S.}~\bibnamefont {Rajendran}},\ }\href
  {\doibase {10.1016/j.physletb.2009.06.011}} {\bibfield  {journal} {\bibinfo
  {journal} {Phys. Lett. B}\ }\textbf {\bibinfo {volume} {678}},\ \bibinfo
  {pages} {37} (\bibinfo {year} {2009})}\BibitemShut {NoStop}%
\bibitem [{\citenamefont {Hohensee}\ \emph {et~al.}(2011)\citenamefont
  {Hohensee}, \citenamefont {Lan}, \citenamefont {Houtz}, \citenamefont {Chan},
  \citenamefont {Estey}, \citenamefont {Kim}, \citenamefont {Kuan},\ and\
  \citenamefont {M{\"u}ller}}]{Hohensee2011}%
  \BibitemOpen
  \bibfield  {author} {\bibinfo {author} {\bibfnamefont {M.}~\bibnamefont
  {Hohensee}}, \bibinfo {author} {\bibfnamefont {S.-Y.}\ \bibnamefont {Lan}},
  \bibinfo {author} {\bibfnamefont {R.}~\bibnamefont {Houtz}}, \bibinfo
  {author} {\bibfnamefont {C.}~\bibnamefont {Chan}}, \bibinfo {author}
  {\bibfnamefont {B.}~\bibnamefont {Estey}}, \bibinfo {author} {\bibfnamefont
  {G.}~\bibnamefont {Kim}}, \bibinfo {author} {\bibfnamefont {P.-C.}\
  \bibnamefont {Kuan}}, \ and\ \bibinfo {author} {\bibfnamefont
  {H.}~\bibnamefont {M{\"u}ller}},\ }\href@noop {} {\bibfield  {journal}
  {\bibinfo  {journal} {Gen. Rel. Gravit}\ }\textbf {\bibinfo {volume} {43}},\
  \bibinfo {pages} {1905} (\bibinfo {year} {2011})}\BibitemShut {NoStop}%
\bibitem [{\citenamefont {Graham}\ \emph {et~al.}(2013)\citenamefont {Graham},
  \citenamefont {Hogan}, \citenamefont {Kasevich},\ and\ \citenamefont
  {Rajendran}}]{Graham2013}%
  \BibitemOpen
  \bibfield  {author} {\bibinfo {author} {\bibfnamefont {P.~W.}\ \bibnamefont
  {Graham}}, \bibinfo {author} {\bibfnamefont {J.~M.}\ \bibnamefont {Hogan}},
  \bibinfo {author} {\bibfnamefont {M.~A.}\ \bibnamefont {Kasevich}}, \ and\
  \bibinfo {author} {\bibfnamefont {S.}~\bibnamefont {Rajendran}},\ }\href
  {\doibase 10.1103/PhysRevLett.110.171102} {\bibfield  {journal} {\bibinfo
  {journal} {Phys. Rev. Lett.}\ }\textbf {\bibinfo {volume} {110}},\ \bibinfo
  {pages} {171102} (\bibinfo {year} {2013})}\BibitemShut {NoStop}%
\bibitem [{\citenamefont {Zhou}\ \emph {et~al.}(2011)\citenamefont {Zhou},
  \citenamefont {Xiong}, \citenamefont {Yang}, \citenamefont {Tang},
  \citenamefont {Peng}, \citenamefont {Hao}, \citenamefont {Li}, \citenamefont
  {Liu}, \citenamefont {Wang},\ and\ \citenamefont {Zhan}}]{Zhou2011}%
  \BibitemOpen
  \bibfield  {author} {\bibinfo {author} {\bibfnamefont {L.}~\bibnamefont
  {Zhou}}, \bibinfo {author} {\bibfnamefont {Z.-Y.}\ \bibnamefont {Xiong}},
  \bibinfo {author} {\bibfnamefont {W.}~\bibnamefont {Yang}}, \bibinfo {author}
  {\bibfnamefont {B.}~\bibnamefont {Tang}}, \bibinfo {author} {\bibfnamefont
  {W.-C.}\ \bibnamefont {Peng}}, \bibinfo {author} {\bibfnamefont
  {K.}~\bibnamefont {Hao}}, \bibinfo {author} {\bibfnamefont {R.-B.}\
  \bibnamefont {Li}}, \bibinfo {author} {\bibfnamefont {M.}~\bibnamefont
  {Liu}}, \bibinfo {author} {\bibfnamefont {J.}~\bibnamefont {Wang}}, \ and\
  \bibinfo {author} {\bibfnamefont {M.-S.}\ \bibnamefont {Zhan}},\ }\href
  {\doibase {10.1007/s10714-011-1167-9}} {\bibfield  {journal} {\bibinfo
  {journal} {Gen. Rel. Gravit}\ }\textbf {\bibinfo {volume} {43}},\ \bibinfo
  {pages} {1931} (\bibinfo {year} {2011})}\BibitemShut {NoStop}%
\bibitem [{\citenamefont {Mazzoni}\ \emph {et~al.}(2015)\citenamefont
  {Mazzoni}, \citenamefont {Zhang}, \citenamefont {Del Aguila}, \citenamefont
  {Salvi}, \citenamefont {Poli},\ and\ \citenamefont {Tino}}]{Mazzoni2015}%
  \BibitemOpen
  \bibfield  {author} {\bibinfo {author} {\bibfnamefont {T.}~\bibnamefont
  {Mazzoni}}, \bibinfo {author} {\bibfnamefont {X.}~\bibnamefont {Zhang}},
  \bibinfo {author} {\bibfnamefont {R.}~\bibnamefont {Del Aguila}}, \bibinfo
  {author} {\bibfnamefont {L.}~\bibnamefont {Salvi}}, \bibinfo {author}
  {\bibfnamefont {N.}~\bibnamefont {Poli}}, \ and\ \bibinfo {author}
  {\bibfnamefont {G.~M.}\ \bibnamefont {Tino}},\ }\href@noop {} {\bibfield
  {journal} {\bibinfo  {journal} {Phys. Rev. A}\ }\textbf {\bibinfo {volume}
  {92}},\ \bibinfo {pages} {053619} (\bibinfo {year} {2015})}\BibitemShut
  {NoStop}%
\bibitem [{\citenamefont {Jamison}\ \emph {et~al.}(2014)\citenamefont
  {Jamison}, \citenamefont {Plotkin-Swing},\ and\ \citenamefont
  {Gupta}}]{Jamison2014}%
  \BibitemOpen
  \bibfield  {author} {\bibinfo {author} {\bibfnamefont {A.~O.}\ \bibnamefont
  {Jamison}}, \bibinfo {author} {\bibfnamefont {B.}~\bibnamefont
  {Plotkin-Swing}}, \ and\ \bibinfo {author} {\bibfnamefont {S.}~\bibnamefont
  {Gupta}},\ }\href {\doibase 10.1103/PhysRevA.90.063606} {\bibfield  {journal}
  {\bibinfo  {journal} {Phys. Rev. A}\ }\textbf {\bibinfo {volume} {90}},\
  \bibinfo {pages} {063606} (\bibinfo {year} {2014})}\BibitemShut {NoStop}%
\bibitem [{\citenamefont {Poli}\ \emph {et~al.}(2011)\citenamefont {Poli},
  \citenamefont {Wang}, \citenamefont {Tarallo}, \citenamefont {Alberti},
  \citenamefont {Prevedelli},\ and\ \citenamefont {Tino}}]{Poli2011}%
  \BibitemOpen
  \bibfield  {author} {\bibinfo {author} {\bibfnamefont {N.}~\bibnamefont
  {Poli}}, \bibinfo {author} {\bibfnamefont {F.-Y.}\ \bibnamefont {Wang}},
  \bibinfo {author} {\bibfnamefont {M.~G.}\ \bibnamefont {Tarallo}}, \bibinfo
  {author} {\bibfnamefont {A.}~\bibnamefont {Alberti}}, \bibinfo {author}
  {\bibfnamefont {M.}~\bibnamefont {Prevedelli}}, \ and\ \bibinfo {author}
  {\bibfnamefont {G.~M.}\ \bibnamefont {Tino}},\ }\href {\doibase
  10.1103/PhysRevLett.106.038501} {\bibfield  {journal} {\bibinfo  {journal}
  {Phys. Rev. Lett.}\ }\textbf {\bibinfo {volume} {106}},\ \bibinfo {pages}
  {038501} (\bibinfo {year} {2011})}\BibitemShut {NoStop}%
\bibitem [{\citenamefont {Tarallo}\ \emph {et~al.}(2014)\citenamefont
  {Tarallo}, \citenamefont {Mazzoni}, \citenamefont {Poli}, \citenamefont
  {Sutyrin}, \citenamefont {Zhang},\ and\ \citenamefont {Tino}}]{Tarallo2014}%
  \BibitemOpen
  \bibfield  {author} {\bibinfo {author} {\bibfnamefont {M.~G.}\ \bibnamefont
  {Tarallo}}, \bibinfo {author} {\bibfnamefont {T.}~\bibnamefont {Mazzoni}},
  \bibinfo {author} {\bibfnamefont {N.}~\bibnamefont {Poli}}, \bibinfo {author}
  {\bibfnamefont {D.~V.}\ \bibnamefont {Sutyrin}}, \bibinfo {author}
  {\bibfnamefont {X.}~\bibnamefont {Zhang}}, \ and\ \bibinfo {author}
  {\bibfnamefont {G.~M.}\ \bibnamefont {Tino}},\ }\href {\doibase
  10.1103/PhysRevLett.113.023005} {\bibfield  {journal} {\bibinfo  {journal}
  {Phys. Rev. Lett.}\ }\textbf {\bibinfo {volume} {113}},\ \bibinfo {pages}
  {023005} (\bibinfo {year} {2014})}\BibitemShut {NoStop}%
\bibitem [{\citenamefont {McDonald}\ \emph {et~al.}(2013)\citenamefont
  {McDonald}, \citenamefont {Keal}, \citenamefont {Altin}, \citenamefont
  {Debs}, \citenamefont {Bennetts}, \citenamefont {Kuhn}, \citenamefont
  {Hardman}, \citenamefont {Johnsson}, \citenamefont {Close},\ and\
  \citenamefont {Robins}}]{McDonald2013}%
  \BibitemOpen
  \bibfield  {author} {\bibinfo {author} {\bibfnamefont {G.~D.}\ \bibnamefont
  {McDonald}}, \bibinfo {author} {\bibfnamefont {H.}~\bibnamefont {Keal}},
  \bibinfo {author} {\bibfnamefont {P.~A.}\ \bibnamefont {Altin}}, \bibinfo
  {author} {\bibfnamefont {J.~E.}\ \bibnamefont {Debs}}, \bibinfo {author}
  {\bibfnamefont {S.}~\bibnamefont {Bennetts}}, \bibinfo {author}
  {\bibfnamefont {C.~C.~N.}\ \bibnamefont {Kuhn}}, \bibinfo {author}
  {\bibfnamefont {K.~S.}\ \bibnamefont {Hardman}}, \bibinfo {author}
  {\bibfnamefont {M.~T.}\ \bibnamefont {Johnsson}}, \bibinfo {author}
  {\bibfnamefont {J.~D.}\ \bibnamefont {Close}}, \ and\ \bibinfo {author}
  {\bibfnamefont {N.~P.}\ \bibnamefont {Robins}},\ }\href {\doibase
  10.1103/PhysRevA.87.013632} {\bibfield  {journal} {\bibinfo  {journal} {Phys.
  Rev. A}\ }\textbf {\bibinfo {volume} {87}},\ \bibinfo {pages} {013632}
  (\bibinfo {year} {2013})}\BibitemShut {NoStop}%
\bibitem [{\citenamefont {Charri\`ere}\ \emph {et~al.}(2012)\citenamefont
  {Charri\`ere}, \citenamefont {Cadoret}, \citenamefont {Zahzam}, \citenamefont
  {Bidel},\ and\ \citenamefont {Bresson}}]{Charriere2012}%
  \BibitemOpen
  \bibfield  {author} {\bibinfo {author} {\bibfnamefont {R.}~\bibnamefont
  {Charri\`ere}}, \bibinfo {author} {\bibfnamefont {M.}~\bibnamefont
  {Cadoret}}, \bibinfo {author} {\bibfnamefont {N.}~\bibnamefont {Zahzam}},
  \bibinfo {author} {\bibfnamefont {Y.}~\bibnamefont {Bidel}}, \ and\ \bibinfo
  {author} {\bibfnamefont {A.}~\bibnamefont {Bresson}},\ }\href {\doibase
  10.1103/PhysRevA.85.013639} {\bibfield  {journal} {\bibinfo  {journal} {Phys.
  Rev. A}\ }\textbf {\bibinfo {volume} {85}},\ \bibinfo {pages} {013639}
  (\bibinfo {year} {2012})}\BibitemShut {NoStop}%
\bibitem [{\citenamefont {Andia}\ \emph {et~al.}(2013)\citenamefont {Andia},
  \citenamefont {Jannin}, \citenamefont {Nez}, \citenamefont {Biraben},
  \citenamefont {Guellati-Kh\'elifa},\ and\ \citenamefont
  {Clad\'e}}]{Andia2013}%
  \BibitemOpen
  \bibfield  {author} {\bibinfo {author} {\bibfnamefont {M.}~\bibnamefont
  {Andia}}, \bibinfo {author} {\bibfnamefont {R.}~\bibnamefont {Jannin}},
  \bibinfo {author} {\bibfnamefont {F.}~\bibnamefont {Nez}}, \bibinfo {author}
  {\bibfnamefont {F.}~\bibnamefont {Biraben}}, \bibinfo {author} {\bibfnamefont
  {S.}~\bibnamefont {Guellati-Kh\'elifa}}, \ and\ \bibinfo {author}
  {\bibfnamefont {P.}~\bibnamefont {Clad\'e}},\ }\href {\doibase
  10.1103/PhysRevA.88.031605} {\bibfield  {journal} {\bibinfo  {journal} {Phys.
  Rev. A}\ }\textbf {\bibinfo {volume} {88}},\ \bibinfo {pages} {031605}
  (\bibinfo {year} {2013})}\BibitemShut {NoStop}%
\bibitem [{\citenamefont {Tarallo}\ \emph {et~al.}(2012)\citenamefont
  {Tarallo}, \citenamefont {Alberti}, \citenamefont {Poli}, \citenamefont
  {Chiofalo}, \citenamefont {Wang},\ and\ \citenamefont {Tino}}]{Tarallo2012}%
  \BibitemOpen
  \bibfield  {author} {\bibinfo {author} {\bibfnamefont {M.~G.}\ \bibnamefont
  {Tarallo}}, \bibinfo {author} {\bibfnamefont {A.}~\bibnamefont {Alberti}},
  \bibinfo {author} {\bibfnamefont {N.}~\bibnamefont {Poli}}, \bibinfo {author}
  {\bibfnamefont {M.~L.}\ \bibnamefont {Chiofalo}}, \bibinfo {author}
  {\bibfnamefont {F.-Y.}\ \bibnamefont {Wang}}, \ and\ \bibinfo {author}
  {\bibfnamefont {G.~M.}\ \bibnamefont {Tino}},\ }\href {\doibase
  10.1103/PhysRevA.86.033615} {\bibfield  {journal} {\bibinfo  {journal} {Phys.
  Rev. A}\ }\textbf {\bibinfo {volume} {86}},\ \bibinfo {pages} {033615}
  (\bibinfo {year} {2012})}\BibitemShut {NoStop}%
\bibitem [{\citenamefont {Szigeti}\ \emph {et~al.}(2012)\citenamefont
  {Szigeti}, \citenamefont {Debs}, \citenamefont {Hope}, \citenamefont
  {Robins},\ and\ \citenamefont {Close}}]{Szigeti2012}%
  \BibitemOpen
  \bibfield  {author} {\bibinfo {author} {\bibfnamefont {S.~S.}\ \bibnamefont
  {Szigeti}}, \bibinfo {author} {\bibfnamefont {J.~E.}\ \bibnamefont {Debs}},
  \bibinfo {author} {\bibfnamefont {J.~J.}\ \bibnamefont {Hope}}, \bibinfo
  {author} {\bibfnamefont {N.~P.}\ \bibnamefont {Robins}}, \ and\ \bibinfo
  {author} {\bibfnamefont {J.~D.}\ \bibnamefont {Close}},\ }\href@noop {}
  {\bibfield  {journal} {\bibinfo  {journal} {New J. Phys.}\ }\textbf {\bibinfo
  {volume} {14}},\ \bibinfo {pages} {023009} (\bibinfo {year}
  {2012})}\BibitemShut {NoStop}%
\bibitem [{\citenamefont {Giltner}\ \emph {et~al.}(1995)\citenamefont
  {Giltner}, \citenamefont {McGowan},\ and\ \citenamefont {Lee}}]{Giltner1995}%
  \BibitemOpen
  \bibfield  {author} {\bibinfo {author} {\bibfnamefont {D.~M.}\ \bibnamefont
  {Giltner}}, \bibinfo {author} {\bibfnamefont {R.~W.}\ \bibnamefont
  {McGowan}}, \ and\ \bibinfo {author} {\bibfnamefont {S.~A.}\ \bibnamefont
  {Lee}},\ }\href {\doibase 10.1103/PhysRevA.52.3966} {\bibfield  {journal}
  {\bibinfo  {journal} {Phys. Rev. A}\ }\textbf {\bibinfo {volume} {52}},\
  \bibinfo {pages} {3966} (\bibinfo {year} {1995})}\BibitemShut {NoStop}%
\bibitem [{\citenamefont {M\"uller}\ \emph {et~al.}(2008)\citenamefont
  {M\"uller}, \citenamefont {Chiow},\ and\ \citenamefont {Chu}}]{Muller2008}%
  \BibitemOpen
  \bibfield  {author} {\bibinfo {author} {\bibfnamefont {H.}~\bibnamefont
  {M\"uller}}, \bibinfo {author} {\bibfnamefont {S.-w.}\ \bibnamefont {Chiow}},
  \ and\ \bibinfo {author} {\bibfnamefont {S.}~\bibnamefont {Chu}},\ }\href
  {\doibase 10.1103/PhysRevA.77.023609} {\bibfield  {journal} {\bibinfo
  {journal} {Phys. Rev. A}\ }\textbf {\bibinfo {volume} {77}},\ \bibinfo
  {pages} {023609} (\bibinfo {year} {2008})}\BibitemShut {NoStop}%
\bibitem [{\citenamefont {Giese}\ \emph {et~al.}(2013)\citenamefont {Giese},
  \citenamefont {Roura}, \citenamefont {Tackmann}, \citenamefont {Rasel},\ and\
  \citenamefont {Schleich}}]{Giese2013}%
  \BibitemOpen
  \bibfield  {author} {\bibinfo {author} {\bibfnamefont {E.}~\bibnamefont
  {Giese}}, \bibinfo {author} {\bibfnamefont {A.}~\bibnamefont {Roura}},
  \bibinfo {author} {\bibfnamefont {G.}~\bibnamefont {Tackmann}}, \bibinfo
  {author} {\bibfnamefont {E.~M.}\ \bibnamefont {Rasel}}, \ and\ \bibinfo
  {author} {\bibfnamefont {W.~P.}\ \bibnamefont {Schleich}},\ }\href {\doibase
  10.1103/PhysRevA.88.053608} {\bibfield  {journal} {\bibinfo  {journal} {Phys.
  Rev. A}\ }\textbf {\bibinfo {volume} {88}},\ \bibinfo {pages} {053608}
  (\bibinfo {year} {2013})}\BibitemShut {NoStop}%
\bibitem [{\citenamefont {Cadoret}\ \emph {et~al.}(2009)\citenamefont
  {Cadoret}, \citenamefont {De~Mirandes}, \citenamefont {Clad{\'e}},
  \citenamefont {Nez}, \citenamefont {Julien}, \citenamefont {Biraben},\ and\
  \citenamefont {Guellati-Kh{\'e}lifa}}]{cadoret2009}%
  \BibitemOpen
  \bibfield  {author} {\bibinfo {author} {\bibfnamefont {M.}~\bibnamefont
  {Cadoret}}, \bibinfo {author} {\bibfnamefont {E.}~\bibnamefont
  {De~Mirandes}}, \bibinfo {author} {\bibfnamefont {P.}~\bibnamefont
  {Clad{\'e}}}, \bibinfo {author} {\bibfnamefont {F.}~\bibnamefont {Nez}},
  \bibinfo {author} {\bibfnamefont {L.}~\bibnamefont {Julien}}, \bibinfo
  {author} {\bibfnamefont {F.}~\bibnamefont {Biraben}}, \ and\ \bibinfo
  {author} {\bibfnamefont {S.}~\bibnamefont {Guellati-Kh{\'e}lifa}},\
  }\href@noop {} {\bibfield  {journal} {\bibinfo  {journal} {Eur. Phys. J.
  Special Topics}\ }\textbf {\bibinfo {volume} {172}},\ \bibinfo {pages} {121}
  (\bibinfo {year} {2009})}\BibitemShut {NoStop}%
\bibitem [{\citenamefont {Andia}\ \emph {et~al.}(2015)\citenamefont {Andia},
  \citenamefont {Wodey}, \citenamefont {Biraben}, \citenamefont {Clad{\'e}},\
  and\ \citenamefont {Guellati-Kh{\'e}lifa}}]{andia2015}%
  \BibitemOpen
  \bibfield  {author} {\bibinfo {author} {\bibfnamefont {M.}~\bibnamefont
  {Andia}}, \bibinfo {author} {\bibfnamefont {{\'E}.}~\bibnamefont {Wodey}},
  \bibinfo {author} {\bibfnamefont {F.}~\bibnamefont {Biraben}}, \bibinfo
  {author} {\bibfnamefont {P.}~\bibnamefont {Clad{\'e}}}, \ and\ \bibinfo
  {author} {\bibfnamefont {S.}~\bibnamefont {Guellati-Kh{\'e}lifa}},\
  }\href@noop {} {\bibfield  {journal} {\bibinfo  {journal} {J. Opt. Soc. Am.
  B}\ }\textbf {\bibinfo {volume} {32}},\ \bibinfo {pages} {1038} (\bibinfo
  {year} {2015})}\BibitemShut {NoStop}%
\bibitem [{\citenamefont {Parazzoli}\ \emph {et~al.}(2012)\citenamefont
  {Parazzoli}, \citenamefont {Hankin},\ and\ \citenamefont
  {Biedermann}}]{Parazzoli2012}%
  \BibitemOpen
  \bibfield  {author} {\bibinfo {author} {\bibfnamefont {L.~P.}\ \bibnamefont
  {Parazzoli}}, \bibinfo {author} {\bibfnamefont {A.~M.}\ \bibnamefont
  {Hankin}}, \ and\ \bibinfo {author} {\bibfnamefont {G.~W.}\ \bibnamefont
  {Biedermann}},\ }\href {\doibase 10.1103/PhysRevLett.109.230401} {\bibfield
  {journal} {\bibinfo  {journal} {Phys. Rev. Lett.}\ }\textbf {\bibinfo
  {volume} {109}},\ \bibinfo {pages} {230401} (\bibinfo {year}
  {2012})}\BibitemShut {NoStop}%
\bibitem [{\citenamefont {Kellogg}\ \emph {et~al.}(2007)\citenamefont
  {Kellogg}, \citenamefont {Yu}, \citenamefont {Kohel}, \citenamefont
  {Thompson}, \citenamefont {Aveline},\ and\ \citenamefont
  {Maleki}}]{kellogg2007}%
  \BibitemOpen
  \bibfield  {author} {\bibinfo {author} {\bibfnamefont {J.~R.}\ \bibnamefont
  {Kellogg}}, \bibinfo {author} {\bibfnamefont {N.}~\bibnamefont {Yu}},
  \bibinfo {author} {\bibfnamefont {J.~M.}\ \bibnamefont {Kohel}}, \bibinfo
  {author} {\bibfnamefont {R.~J.}\ \bibnamefont {Thompson}}, \bibinfo {author}
  {\bibfnamefont {D.~C.}\ \bibnamefont {Aveline}}, \ and\ \bibinfo {author}
  {\bibfnamefont {L.}~\bibnamefont {Maleki}},\ }\href@noop {} {\bibfield
  {journal} {\bibinfo  {journal} {J. Mod. Opt}\ }\textbf {\bibinfo {volume}
  {54}},\ \bibinfo {pages} {2533} (\bibinfo {year} {2007})}\BibitemShut
  {NoStop}%
\bibitem [{\citenamefont {Charri{\`e}re}(2011)}]{charriere2011}%
  \BibitemOpen
  \bibfield  {author} {\bibinfo {author} {\bibfnamefont {R.}~\bibnamefont
  {Charri{\`e}re}},\ }\emph {\bibinfo {title} {Optimisation d'un capteur
  inertiel {\`a} atomes froids par une nouvelle technique de mesure
  acc{\'e}l{\'e}rom{\'e}trique combinant interf{\'e}rom{\'e}trie atomique et
  oscillations de Bloch}},\ \href@noop {} {Ph.D. thesis},\ \bibinfo  {school}
  {Paris 6} (\bibinfo {year} {2011})\BibitemShut {NoStop}%
\bibitem [{\citenamefont {Hamilton}\ \emph {et~al.}(2015)\citenamefont
  {Hamilton}, \citenamefont {Jaffe}, \citenamefont {Brown}, \citenamefont
  {Maisenbacher}, \citenamefont {Estey},\ and\ \citenamefont
  {M\"uller}}]{Hamilton2015}%
  \BibitemOpen
  \bibfield  {author} {\bibinfo {author} {\bibfnamefont {P.}~\bibnamefont
  {Hamilton}}, \bibinfo {author} {\bibfnamefont {M.}~\bibnamefont {Jaffe}},
  \bibinfo {author} {\bibfnamefont {J.~M.}\ \bibnamefont {Brown}}, \bibinfo
  {author} {\bibfnamefont {L.}~\bibnamefont {Maisenbacher}}, \bibinfo {author}
  {\bibfnamefont {B.}~\bibnamefont {Estey}}, \ and\ \bibinfo {author}
  {\bibfnamefont {H.}~\bibnamefont {M\"uller}},\ }\href {\doibase
  10.1103/PhysRevLett.114.100405} {\bibfield  {journal} {\bibinfo  {journal}
  {Phys. Rev. Lett.}\ }\textbf {\bibinfo {volume} {114}},\ \bibinfo {pages}
  {100405} (\bibinfo {year} {2015})}\BibitemShut {NoStop}%
\end{thebibliography}
\end{document}